\documentclass[11pt,a4paper]{article}

\pdfoutput=1

\usepackage{jheppub}
\usepackage{amsmath}
\usepackage{amsfonts}
\usepackage{graphicx}
\usepackage{amssymb}
\usepackage{amssymb}
\usepackage{latexsym}
\usepackage{array}
\usepackage{bbold}
\usepackage[normalem]{ulem}
\usepackage{xcolor}
\usepackage{slashed}
\usepackage{cancel}

\newcommand{\bea}{\begin{eqnarray}}
\newcommand{\eea}{\end{eqnarray}}
\newcommand{\be}{\begin{equation}}
\newcommand{\ee}{\end{equation}}

\usepackage{xcolor}
\definecolor{mygreen}{HTML}{006E28}

\title{Excited oscillons and charge-swapping}

\author[a,b]{A. Alonso-Izquierdo}
\affiliation[a]{Departamento de Matem\'atica Aplicada, University of Salamanca, Casas del Parque 2, 37008 -
Salamanca, Spain}
\affiliation[b]{IUFFyM, University of Salamanca, Plaza de la Merced 1, 37008 - Salamanca, Spain}
\emailAdd{alonsoiz@usal.es}

\author[b]{D. Canillas Mart\'inez}
\emailAdd{dnl.canillas@usal.es}

\author[c]{T. Roma\'nczukiewicz}
\emailAdd{tomasz.romanczukiewicz@uj.edu.pl}

\author[c]{K. S\l awi\'nska}
\affiliation[c]{Faculty of Theoretical Physics, Astronomy and Applied Computer Science, Jagiellonian University, Krak\'ow, Poland}
\emailAdd{katarzyna.slawinska@uj.edu.pl}

\author[c,d]{A. Wereszczy\'nski}
\affiliation[d]{International Institute for Sustainability with Knotted Chiral Meta Matter (WPI-SKCM2), Higashi-Hiroshima, 1-3-1 Kagamiyama, Hiroshima 739-8526, Japan}
\emailAdd{andrzej.wereszczynski@uj.edu.pl}

\abstract{We show that the charge-swapping phenomenon can be understood as a real valued oscillon carrying an excitation in imaginary direction in the target space. Furthermore, we introduce a two dimensional collective model which quantitatively captures the charge swapping dynamics.
}

\begin{document}

\maketitle

\section{Introduction}
$Q$-balls are non-topological solitons, that are localized, particle-like, nonperturbative solutions, which carry a $U(1)$ Noether charge \cite{Friedberg:1976me, Coleman:1985ki}. This charge arises from the global phase invariance of the action and therefore requires at least a complex field $\phi$. In contrast to a topological charge, this charge is not quantized. 

There are many areas of physics where $Q$-balls found an application. In cosmology, they are expected to be produced in the evolution of the early universe. In addition, they are considered as candidates for dark matter \cite{Kusenko:1997si,Dine:2003ax, Enqvist:2003gh, Cardoso:2019rvt}. After coupling with gravity, they also give rise to boson stars \cite{PhysRev.172.1331, PhysRev.187.1767}, i.e., hypothetical compact objects, whose imprint on gravitational wave measurements is now of great interest; see \cite{Liebling:2012fv} for a review. The $Q$-balls can also be realized in condensed matter systems \cite{Enqvist:2003zb}, e.g., in a superfluid \cite{Bunkov:2007fe}. 

The simplest $Q$-ball is a stationary solution, rotating in the complex target space with a constant frequency. Their properties, stability, and structure of small perturbations are quite well understood. If placed together, $Q$-balls begin to interact \cite{Battye:2000qj,Bowcock, Axenides:1999hs}. In some cases, this leads to the so-called {\it charge-swapping} phenomenon \cite{Copeland:2014qra}. This is a quasi-stable solution in which the $U(1)$ charge periodically flows between two or more spatial regions. 

Typically, charge-swapping solutions are found from an initial condition representing a pair of separated $Q$-ball and anti-$Q$-ball \cite{Copeland:2014qra, Xie:2021glp}. This led to an interpretation of the charge-swapping phenomenon as a $QQ^*$ bound state, where the constituent solitons form a kind of molecule \cite{Zhou:2024mea}. More complicated examples, arising from a larger number of initial $QQ^*$ pairs, were also presented. 

\newpage

In the current work, we present arguments that such an interpretation is not valid. Surprisingly, despite the fact that it can be generated from $QQ^*$ initial data, the charge-swapping solution should be treated as an {\it excited oscillon}. (Concretely, this is an oscillon perturbed along the perpendicular direction in the target space.) This is another type of localized non-perturbative objects, whose stability does not rely on any topological or non-topological charge but is an effect of nonlinear self-interaction \cite{Bogolyubsky:1976nx, Gleiser:1993pt, Copeland:1995fq}. In fact, ocillons are not ultimately stable objects. All the time, they lose energy through radiation \cite{Fodor:2008du, Fodor:2009kf} and eventually decay to vacuum. Nevertheless, the life-time of an oscillon can be extremely long \cite{Graham:2006xs, Salmi:2012ta, Zhang:2020bec, Olle:2020qqy, vanDissel:2023zva}, sometimes comparable to the age of the universe, which makes them important in the dynamics of various solitons. Similarly to $Q$-balls they have various cosmological and astrophysical applications, for example \cite{Gleiser:2011xj, Amin:2011hj, Zhou:2013tsa, Olle:2019kbo, Aurrekoetxea:2023jwd}. 

The main support for our conclusion is found in the mode structure of the charge-swapping states, which is clearly inherited from the oscillon, not from the $Q$-balls. Furthermore, charge-swapping solutions are generated from much more general initial data, which represent such a perturbed oscillon rather than only a $QQ^*$ pair.
 
\section{The complex $\phi^6$ model}
\subsection{The $Q$-balls}
In this work we will focus on the simplest $Q$-ball theory which is the complex $\phi^6$ model in (1+1) dimensions
\begin{equation}
    \mathcal{L} = \partial_\mu\phi\partial^\mu\phi^*-|\phi|^2-|\phi|^4+\beta |\phi|^6\,.
    \label{lag}
\end{equation}
The equations of motion are
\begin{equation}
\phi_{tt}-\phi_{xx}+(1+2|\phi|^2-3\beta |\phi|^4) \phi=0
\label{field_eq}
\end{equation}
and the complex conjugation. Due to the global $U(1)$ symmetry, $\phi \to e^{i \lambda}\phi$,  there is a conserved current $j^\mu$ and  conserved Noether charge $Q$
\begin{equation}
    j_\mu =i \left( \phi^* \partial_\mu \phi -  \phi \partial_\mu \phi^* \right), \;\;\ Q=\int_{-\infty}^\infty \rho \, dx. 
\end{equation}
Here $j^0\equiv \rho$. The simplest, single Q-ball solution is a stationary configuration rotating in the target space with the frequency $\omega_0$
\begin{equation}
   \phi= f_{\omega_0}(x) e^{i\omega_0 t}, 
\label{Q-solution}
\end{equation}
which profile $f_{\omega_0}(x)$ reads
\begin{equation}
  f_{\omega_0}(x)=\frac{\sqrt{2}\epsilon }{\sqrt{1+\sqrt{1-4\beta\epsilon ^2}\cosh(2\epsilon x)}}\,.
\label{solution}
\end{equation}
Here, $\epsilon =\sqrt{1-\omega_0^2}$. The $Q$-ball carries the following energy and the $U(1)$ charge
\begin{equation}
    E (\omega_0) = \frac{4\omega_0 {\epsilon } + Q (4\beta-1 + 4 \beta \omega_0^2 )}{8\omega_0 \beta},\;\;\;\; 
Q(\omega_0) = \frac{4\omega_0}{\sqrt \beta}{\rm arctanh}\left(\frac{1-\sqrt{1-4\beta {\epsilon }^2}}{2{\epsilon } \sqrt \beta}\right)\,.
\label{EQ}
\end{equation}
Although the $Q$-balls exist for $\omega_0 \in [ \omega_{min}, \omega_{max}]$, where $\omega^2_{min}=1-\frac{1}{4\beta}$ and $\omega_{max}=1$, they are classically stable only if 
\begin{equation}
    \frac{\omega_0}{Q} \frac{dQ}{d\omega_0} <0. 
\end{equation}

It has been recently noticed that $Q$-balls possess quite a rich structure of linear modes  \cite{Ciurla:2024ksm}
\begin{equation}
    \delta \phi (x,t) = \eta_1(x)e^{i(\omega_0+\rho)t} +\eta_2(x)e^{i(\omega_0-\rho)t}.
\end{equation}
Note that a mode has two components $\eta_{1,2}$ with frequencies $\omega_0\pm \rho$. Therefore, for stable $Q$-balls, three possibilities can be identified. There can be a bound mode (BM) if the frequencies of both components are below the mass threshold. If one frequency is above the mass threshold, we have a quasinormal mode (QNM) known as the Feshbach resonance. In this case, one component of the mode is still bounded to the $Q$-ball. If both frequencies are above the mass threshold, we get a genuine scattering mode. Of course, there are also zero modes that reflect translational and $U(1)$ symmetries. In the unstable regime, an unstable mode additionally appears. 

\begin{figure}
    \centering
    \includegraphics[width=1\linewidth]{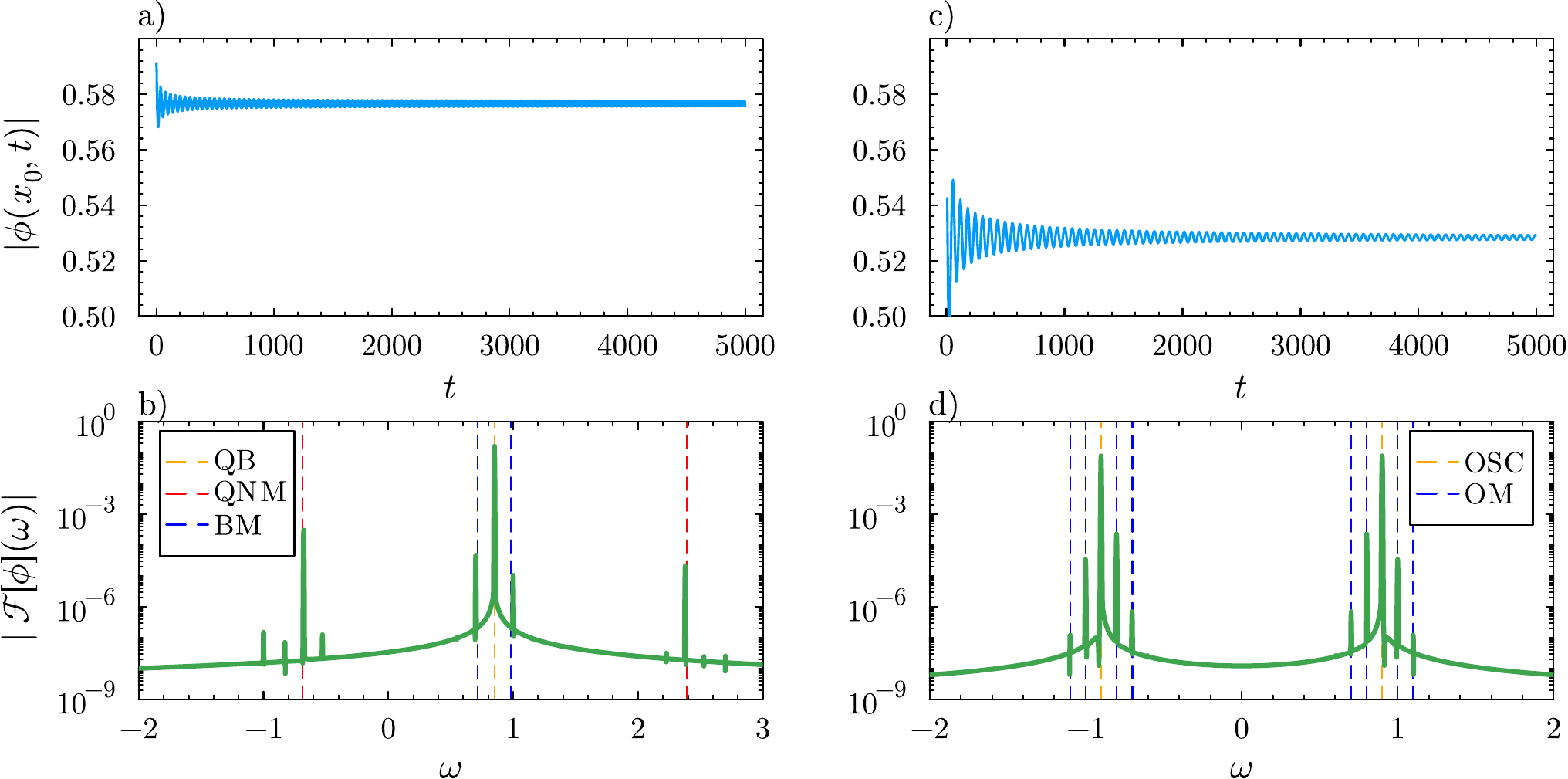}
    \caption{Time dependence of the maximum of Re $\phi$ and the power spectrum of a perturbed $Q$-ball with $\omega_0=0.85$ and $\lambda=1.05$ (left) and oscillon (right). Here $\beta=0.5$.}
    \label{fig:QB-modes}
\end{figure}

In Fig. \ref{fig:QB-modes}, left panels, we display the mode structure of a perturbed $Q$-ball. Specifically, we take $\beta=0.5$ and $\omega_0=0.85$. Perturbation is provided by a radial squeeze by factor $\lambda=1.05$. In the spectrum, there is the main frequency (QB) at $\omega_0$, a bound mode (BM) with $\rho=0.133612$ and a quasinormal mode (QNM) with $\rho=1.538789 + 1.18\cdot 10^{-5} i$, highlighted respectively by orange, blue and red dashed lines. 

\subsection{The Oscillons}
The (complex) $\phi^6$ model, as many models with nonlinear self-interaction, supports oscillons, i.e., approximately periodic in time, localized field configurations. 
Recently, it has been shown that the oscillon and $Q$ balls are, in fact, very closely related to each other \cite{Blaschke:2024dlt}. Especially in the $\phi^6$ theory, these objects are somehow dual to each other \cite{Blaschke:2025anm}. An oscillon can be viewed as formed from two opposite charge $Q$-balls or, vice versa, a $Q$-ball can be treated as a bound state of two oscillons, each in each real and imaginary component of the complex field \cite{Copeland:2014qra, Blaschke:2025anm}. In addition, also their mode structures are related \cite{Blaschke:2025anm}.

In Fig. \ref{fig:QB-modes}, right panel, we show an example of the oscillon generated from the initial profile $\phi(x,0)= \frac{2}{\sqrt{3}}\frac{\epsilon}{\cosh (\epsilon x)}$. The spectrum is $\mathbb{Z}_2$ symmetric. The main peak at $\omega_{O}$ corresponds to the fundamental frequency of the oscillon (OSC). The other peaks at $\omega_O\pm n\omega_{mod}$ (OM) can be interpreted as an effect of the appearance of another small-amplitude oscillon. Importantly, $\omega_{mod}$ is the frequency of the amplitude modulation. Here, $\omega_O=0.90$ and $\omega_{mod}=0.10$.

\section{Charge-swapping}
\subsection{Charge-swapping from $QQ^*$ Initial Conditions}
A charge-swapping solution is typically generated from the initial state being a separated pair of the $Q$-ball and anti-$Q$-ball \cite{Copeland:2014qra, Xie:2021glp}
\begin{equation}
    \Phi(x,t)=f_{\omega_0}(x+x_0) e^{i\omega_0 t} + f_{\omega_0}(x-x_0) e^{-i\omega_0 t}. \label{QQ*-ini}
\end{equation}
Depending on the separation $d=2x_0$ this configuration relaxes to two excited $Q$-balls or to a charge swapping state. 

In Fig. \ref{fig:CS-exampl} we present examples of dynamics generated from the initial data (\ref{QQ*-ini}). Here we fix $\beta=0.5$ and $\omega_0=0.85$. We plot the time evolution of the charge density as well as the real and imaginary components of the complex field. In the upper panels, we show the case where the $Q$-balls remain independent. In the middle panel, there is a single-bounce solution, while in the bottom panel, there is a dynamics with charge swapping. 

\begin{figure}
    \centering
    \includegraphics[width=0.32\linewidth]{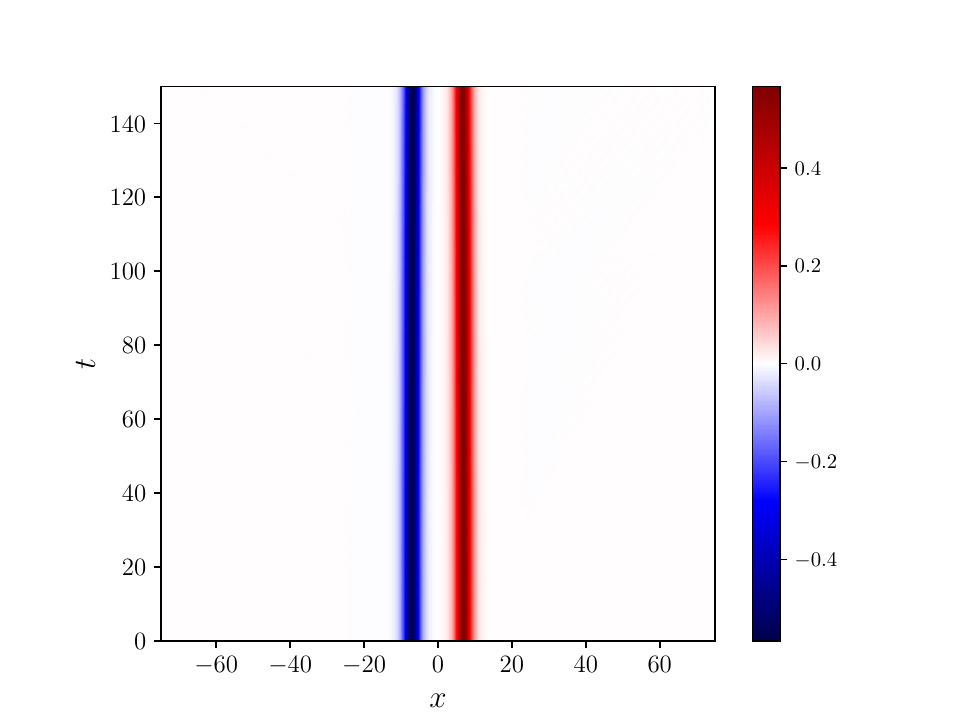}
    \includegraphics[width=0.32\linewidth]{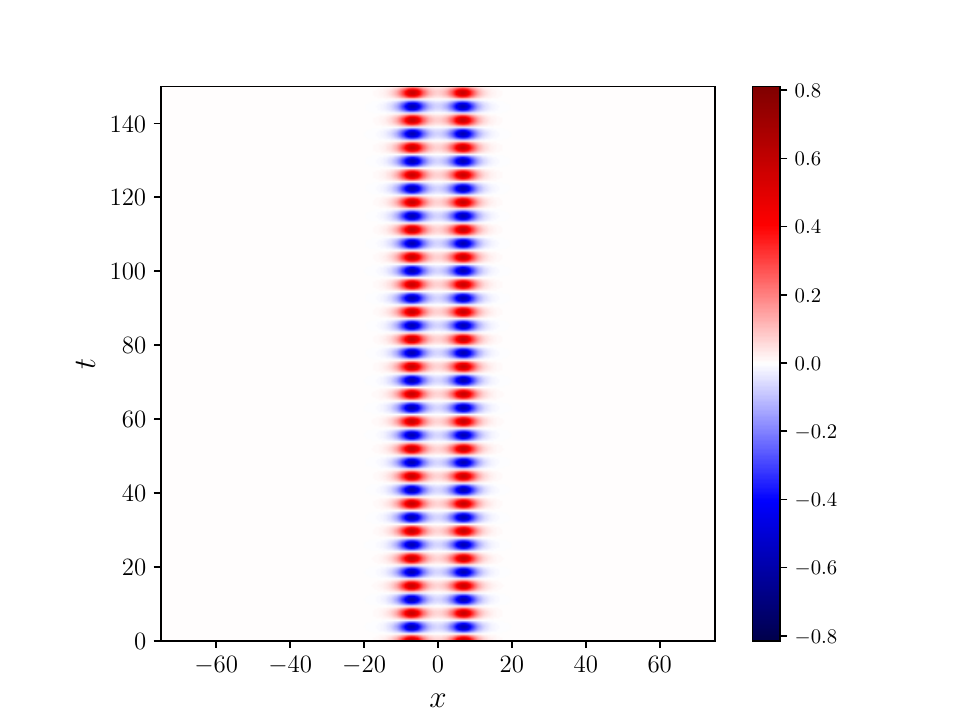}
    \includegraphics[width=0.32\linewidth]{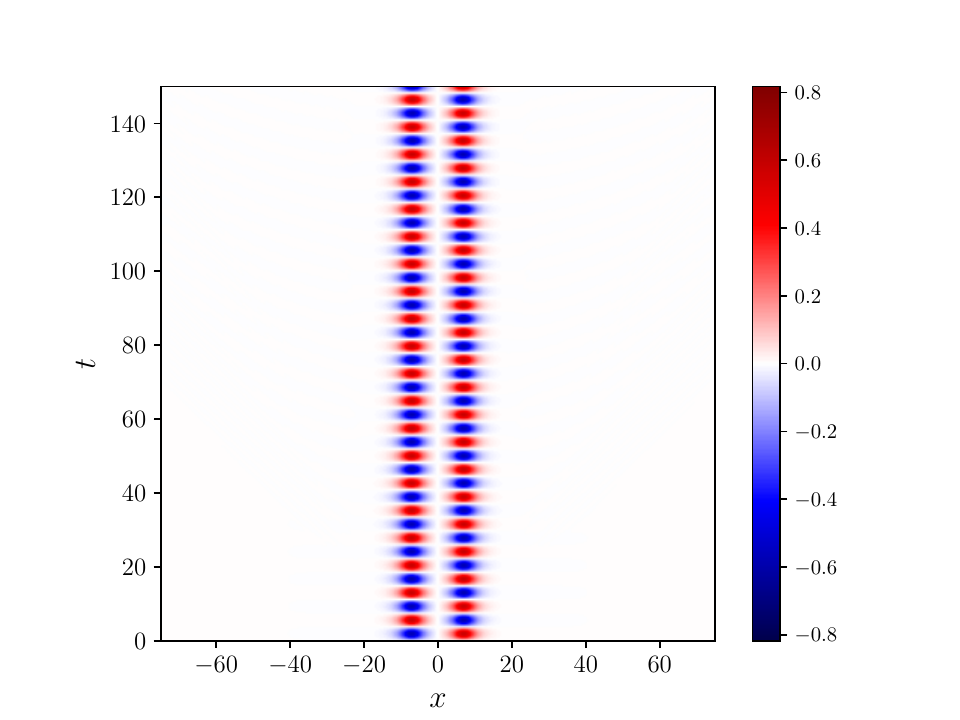}
    \includegraphics[width=0.32\linewidth]{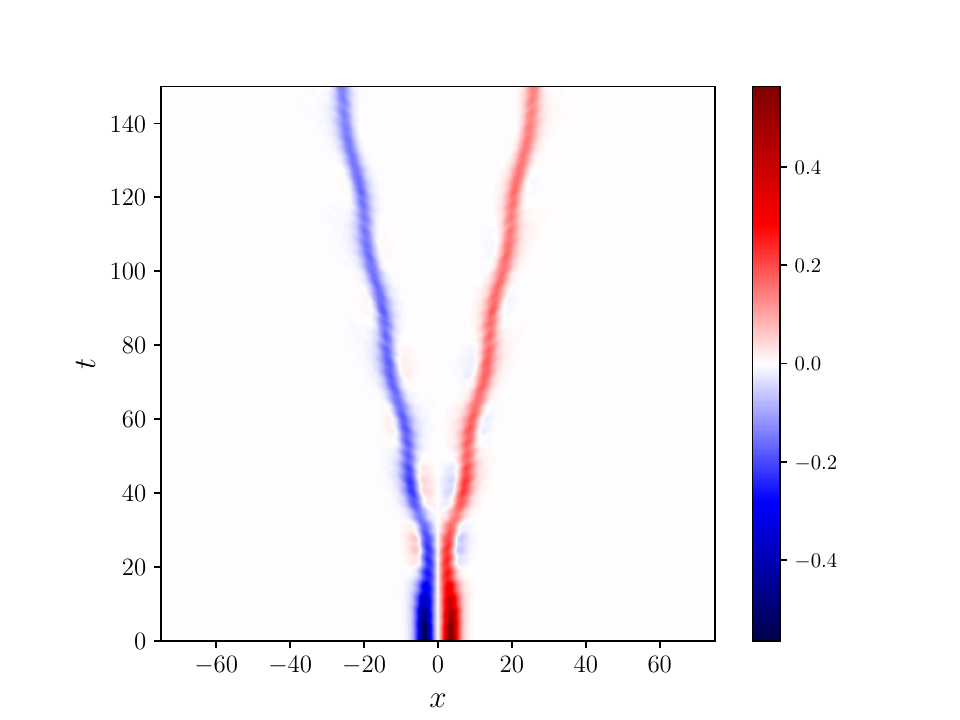}
    \includegraphics[width=0.32\linewidth]{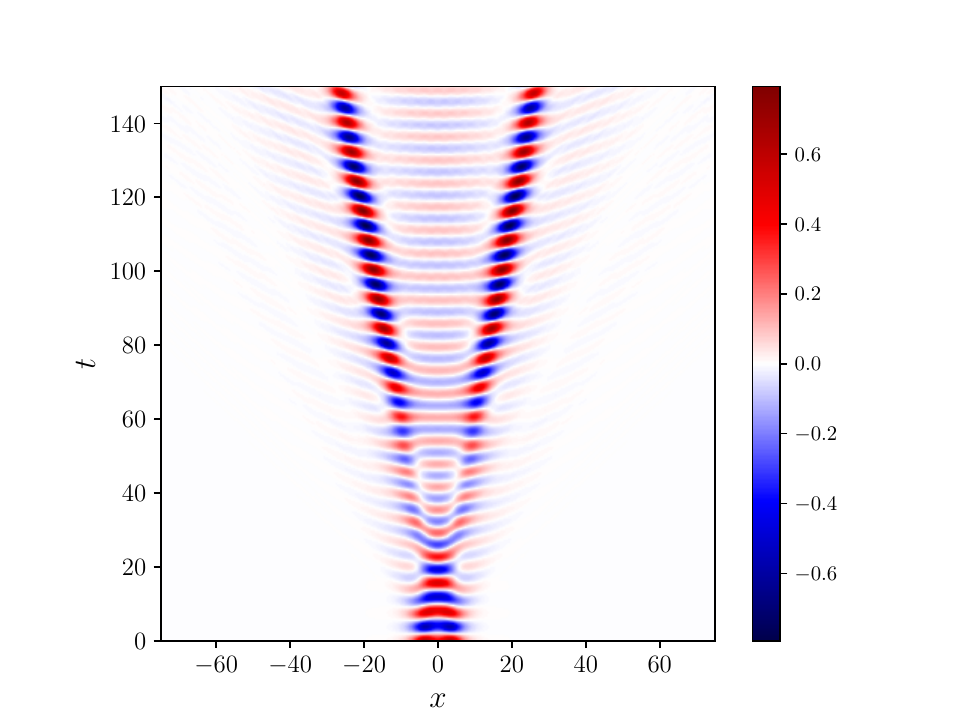}
    \includegraphics[width=0.32\linewidth]{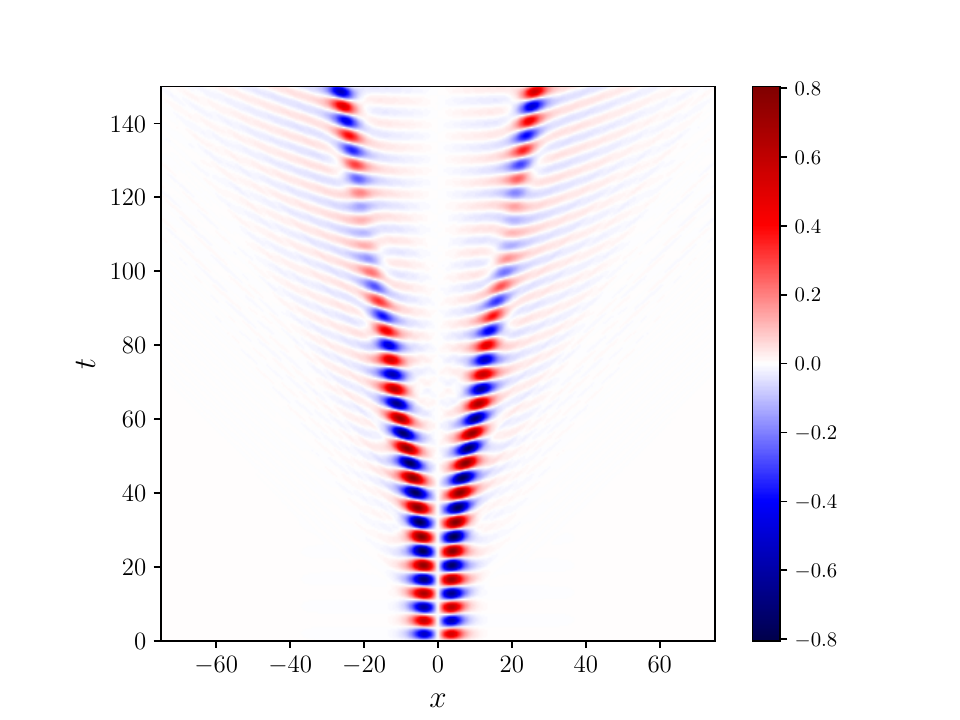}
    \includegraphics[width=0.32\linewidth]{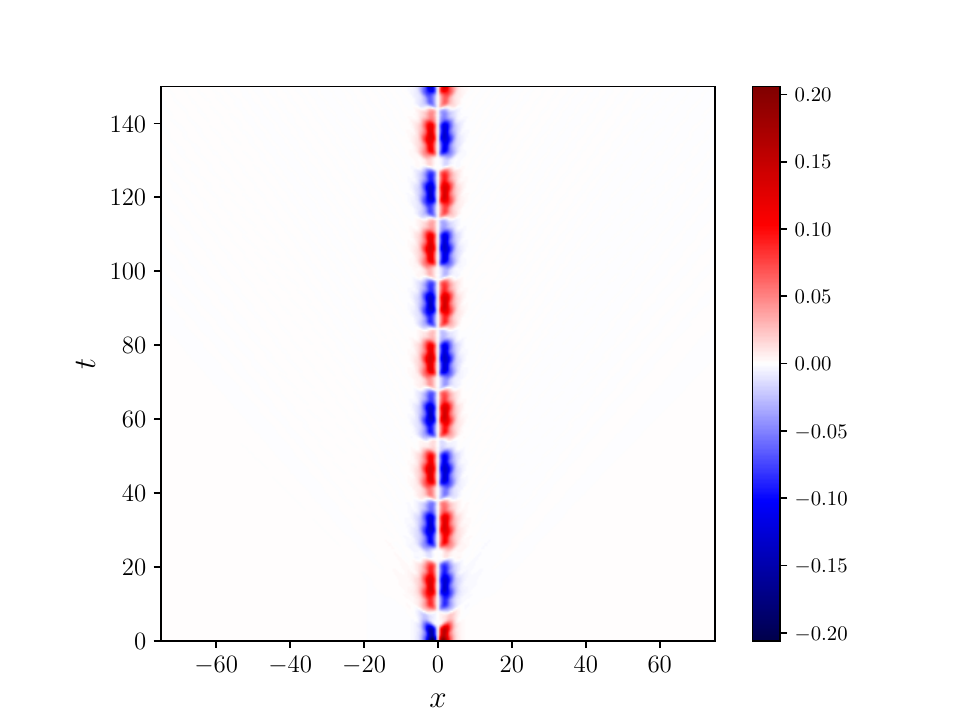}
    \includegraphics[width=0.32\linewidth]{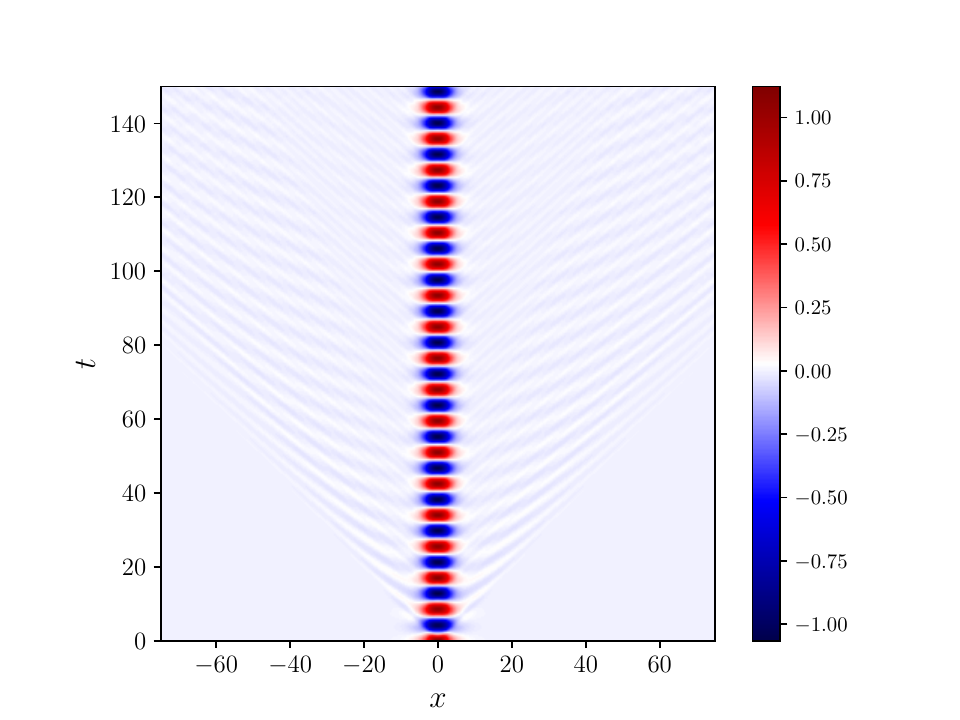}
    \includegraphics[width=0.32\linewidth]{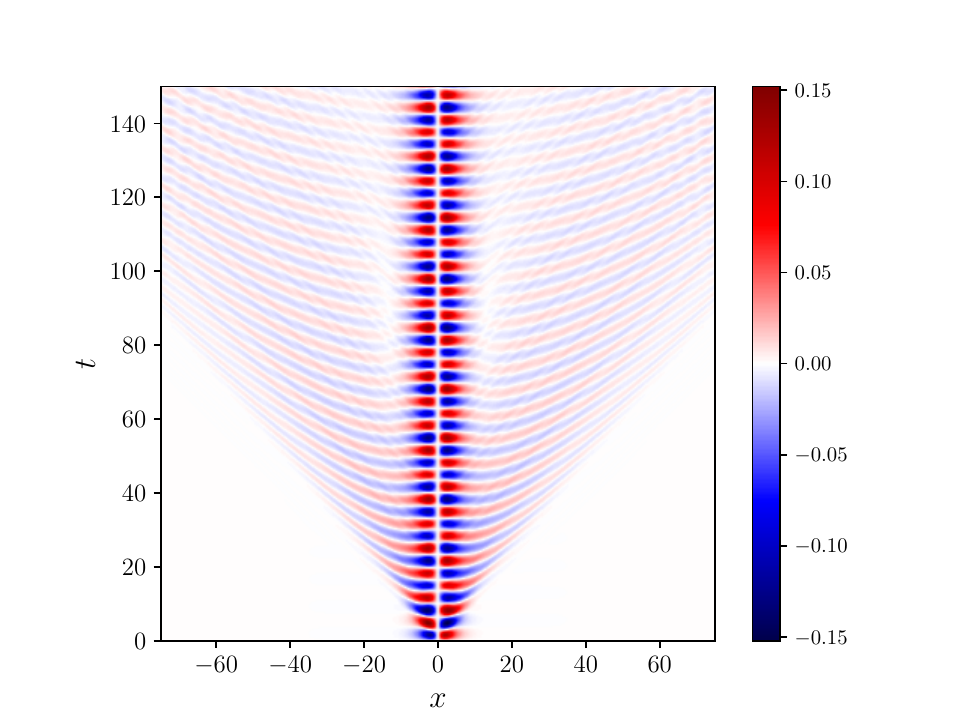}
    \caption{Examples of time evolution obtained from $Q$-ball-anti-$Q$-ball initial data (\ref{QQ*-ini}). Here $\beta=0.5$ and $\omega_0=0.85$ while $x_0=7.5$ (upper row), $x_0=3.5$ (middle row) and $x_0=0.5$ (lower row). We plot: the charge density (left column), Re $\phi$ (middle column) and Im $\phi$ (right column). }
    \label{fig:CS-exampl}
\end{figure}

\begin{figure}[!h] 
    \centering
    \includegraphics[width=0.75\linewidth]{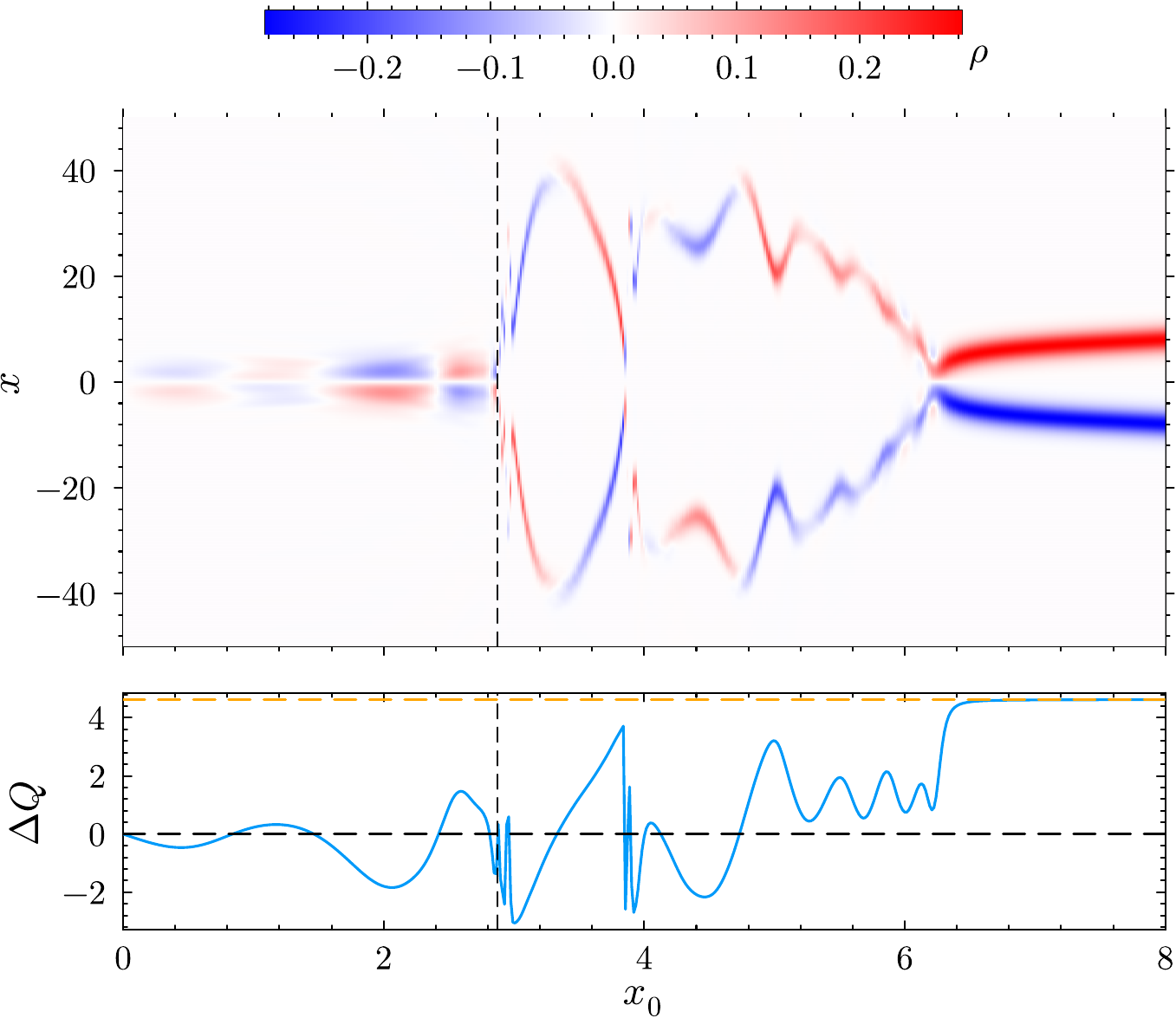}
    \includegraphics[width=0.75\linewidth]{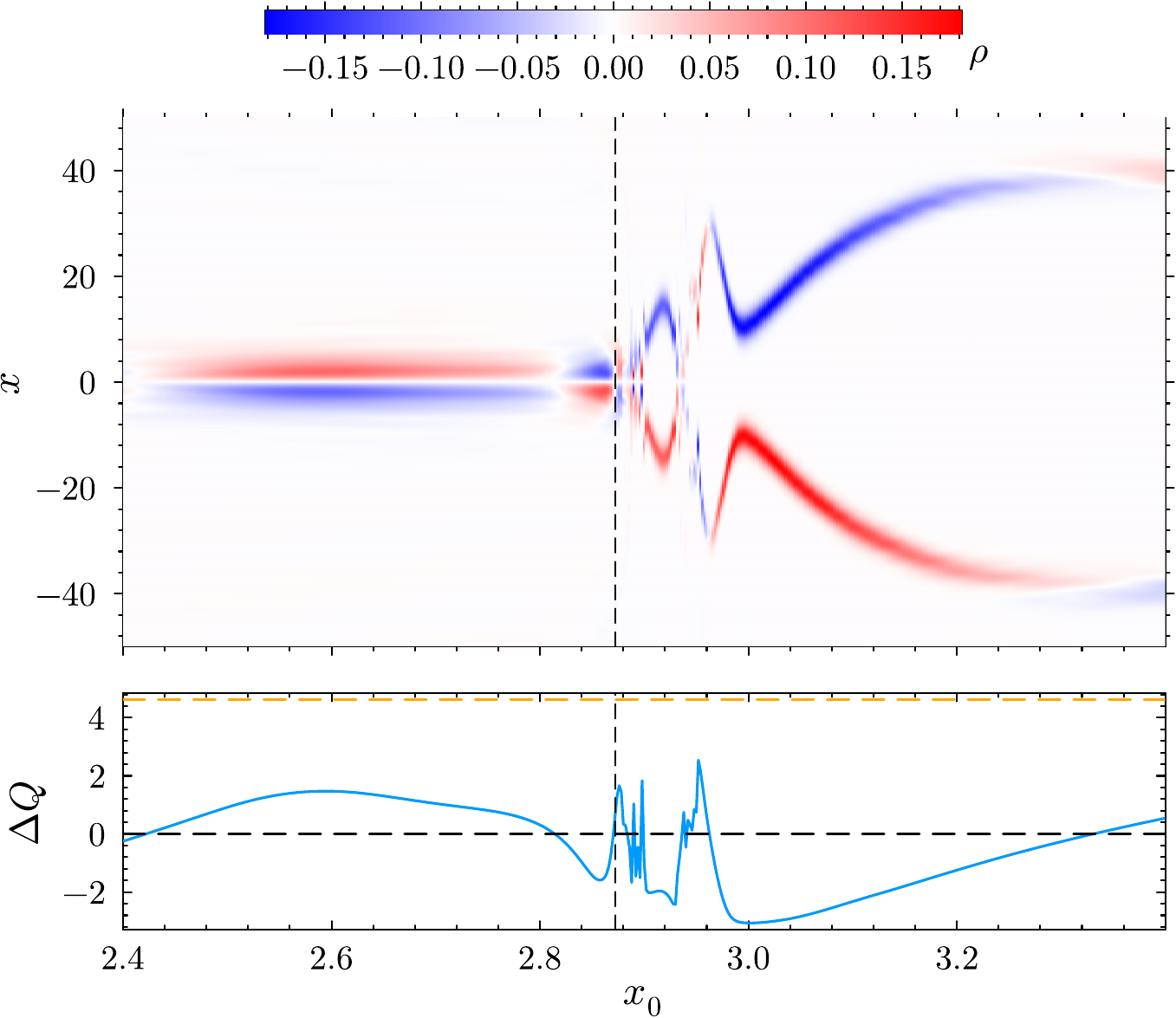}
    \caption{The charge density at $t_{max}=300$ as a function of parameter $x_0$. The vertical line at $x_{cr}=2.87(2)$ separates the charge-swapping regime from other scenarios: the multi-bounce solutions and separated $Q$-balls. Here $\Delta Q = Q(x<0) - Q(x>0)$.}
    \label{fig:x0-scan}
\end{figure}

In general, we may identify three regimes. If the initial separation is large enough, $x_0>\tilde{x}$, the force between the $Q$ balls is weak, and they remain independent throughout the simulation, which in our case ends at $t_{max}=500$. For smaller separations, $\tilde{x} > x_0 >x_{cr}$ the attraction becomes stronger, and $Q$-balls collide. They bounce once and separate or enter into a rather irregular, chaotic multi-bounce regime. It is clearly seen that after the first collision the bouncing $Q$-balls vibrate. Eventually, for even smaller initial separation, $x_0<x_{cr}$, the $QQ^*$ initial data  form a well-visible and quite stable charge-swapping solution. The critical value $x_{cr}$ depends on the model parameter $\beta$ and the parameter $\omega_0$ of the initial state. On the other hand, $\tilde{x}$ is not a universal quantity. For the growing time of the simulation $\tilde{x}$ also increases. This is because there is always an attractive force between the $Q$-ball and the anti-$Q$-ball. Hence, eventually the solitons will always collide.

\begin{figure}
    \centering
    \includegraphics[width=0.8\linewidth]{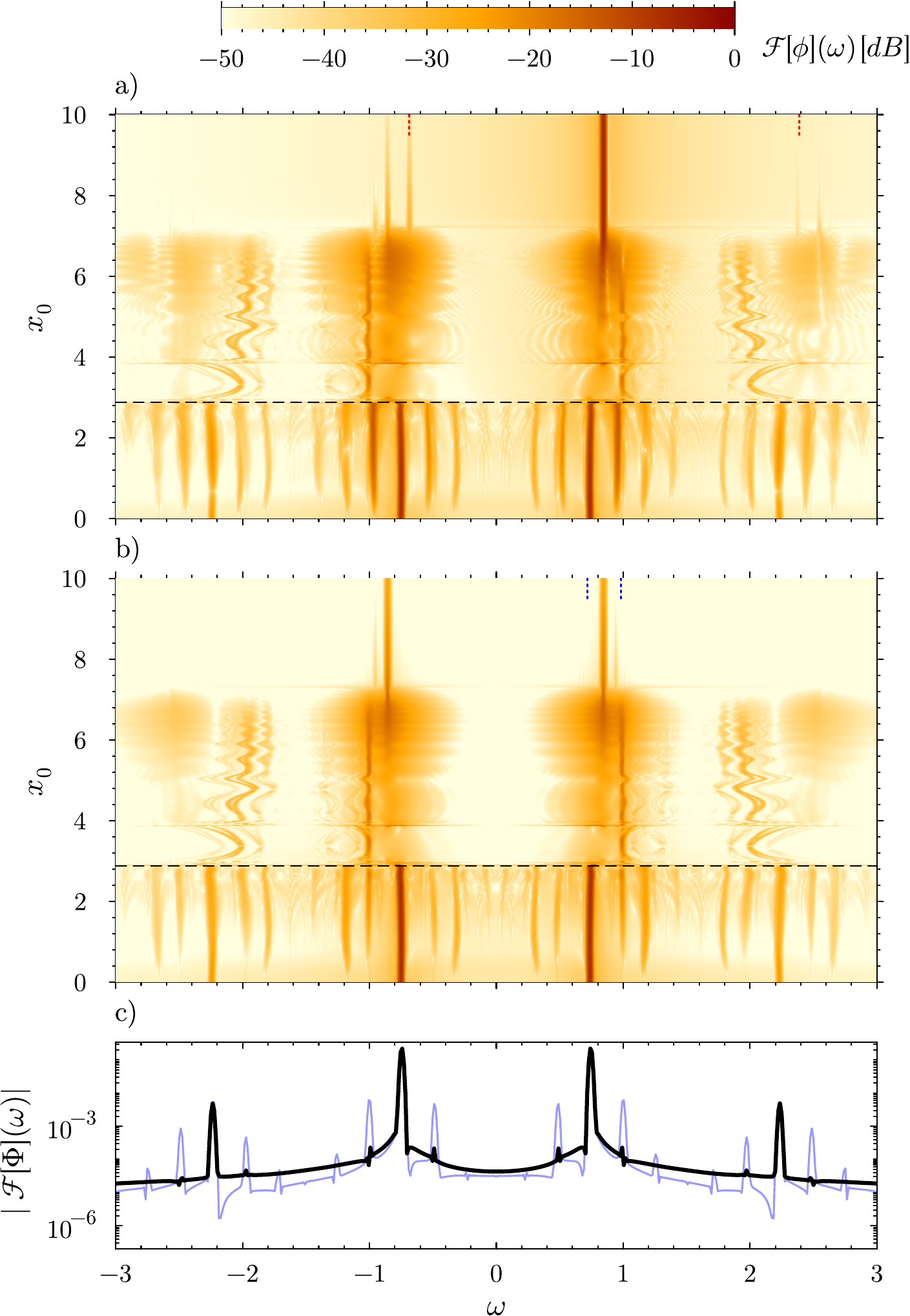}
    \caption{Power spectrum at $x=x_0$ (upper panel) and $x=0$ (lower panel) for the solution found from the $QQ^*$ initial configuration (\ref{QQ*-ini}). Here $\beta=0.5$, $\omega_0=0.85$ and time of the evolution is $t_{max}=500$. At the bottom the power spectrum of the oscillon (blue) and an excited oscillon (black). Horizontal dashed line denotes $x_{cr}=2.87(2)$.}
    \label{fig:CS-scan}
\end{figure}

In Fig. \ref{fig:x0-scan}, upper panel, we present the dependence of the final state of the evolution on the value of the parameter $x_0$ in the initial configuration. We clearly see the three regimes mentioned above. Concretely, we plot the charge distribution at $t_{max}=300$ for $\beta=0.5$ and $\omega_0=0.85$. The critical value of the parameter $x_0$ is $x_{cr}=2.87(2)$ while $\tilde{x}\approx 6.1$. Interestingly, the multi-bounce regime reveals a self-similar pattern which suggests chaotic behaviour, see Fig. \ref{fig:x0-scan} lower panel, where we zoom on the previous plot. This fractal structure occurs in various solitonic processes, such as, for example, kink-antikink scatterings \cite{Campbell:1983xu, Sugiyama:1979mi, Manton:2021ipk} (also \cite{Takyi:2016tnc, Lima:2021jxl, Campos:2021mkn}), collisions of vortices \cite{Krusch:2024vuy, AlonsoIzquierdo:2024nbn} or oscillon decay \cite{Blaschke:2024uec}, and is triggered by the resonance transfer mechanism. Furthermore, in this plot, we also show how the difference between the charges stored in $x>0$ and $x<0$ evolves in time, $\Delta Q =  Q(x<0) - Q(x>0)$, where $Q(x<0)$  is the charge in half space with $x<0$.

\begin{figure}
    \centering
     \includegraphics[width=0.48\linewidth]{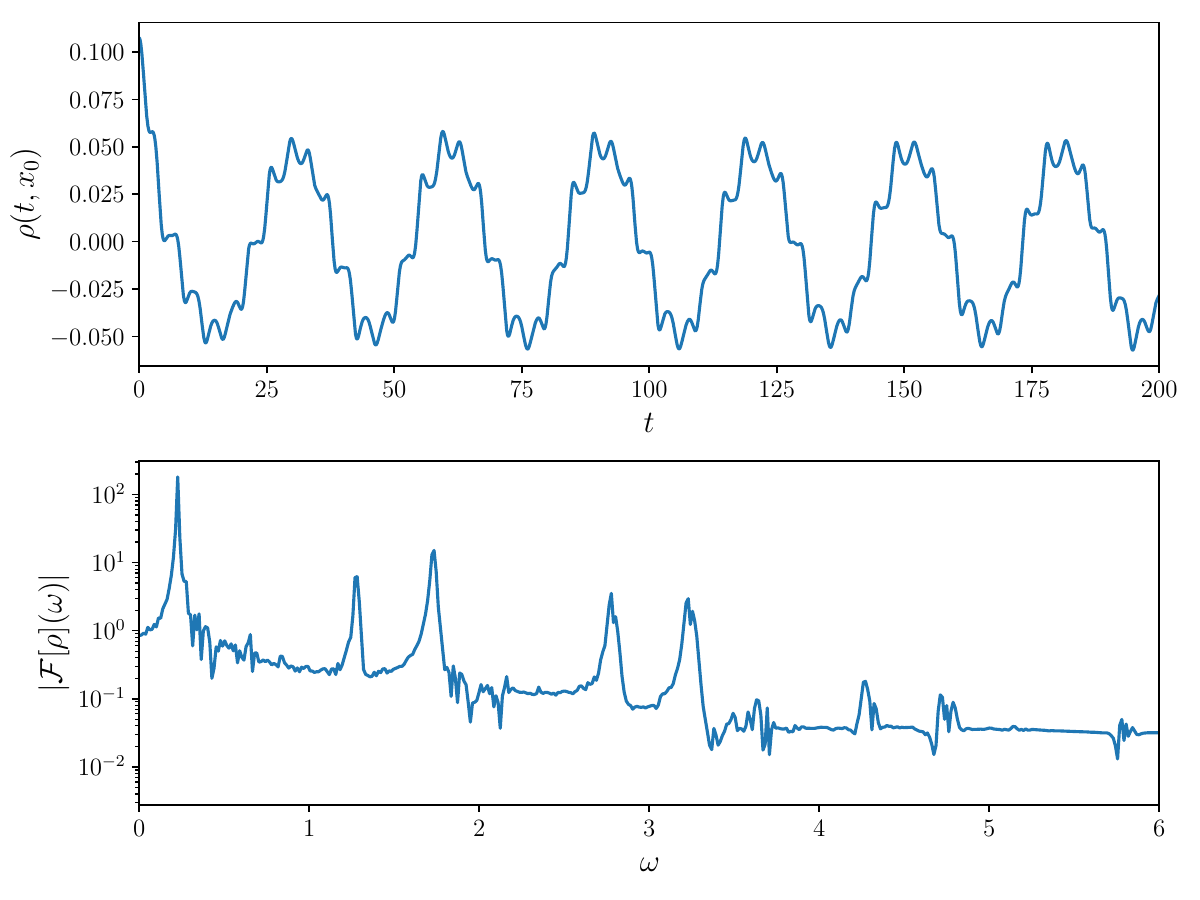}
    \includegraphics[width=0.48\linewidth]{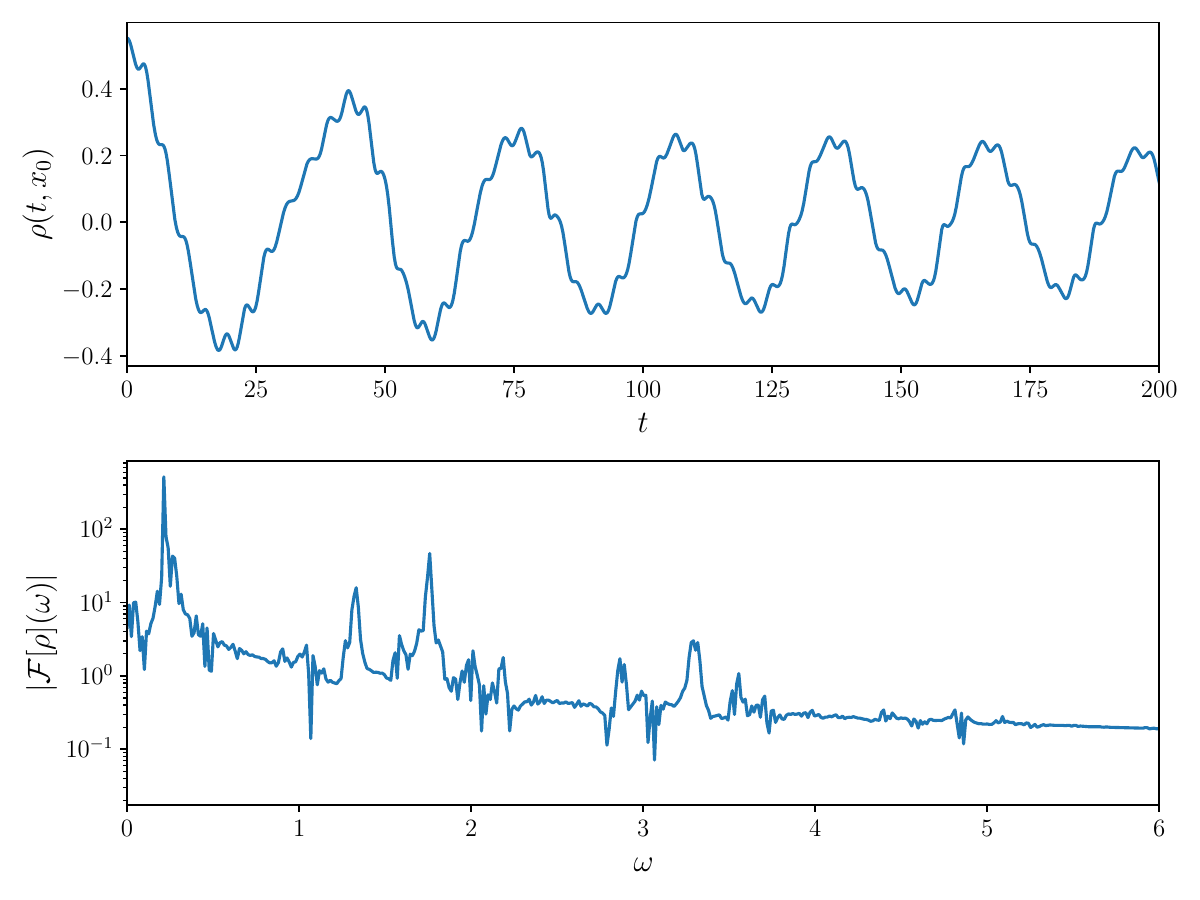}
    \includegraphics[width=0.48\linewidth]{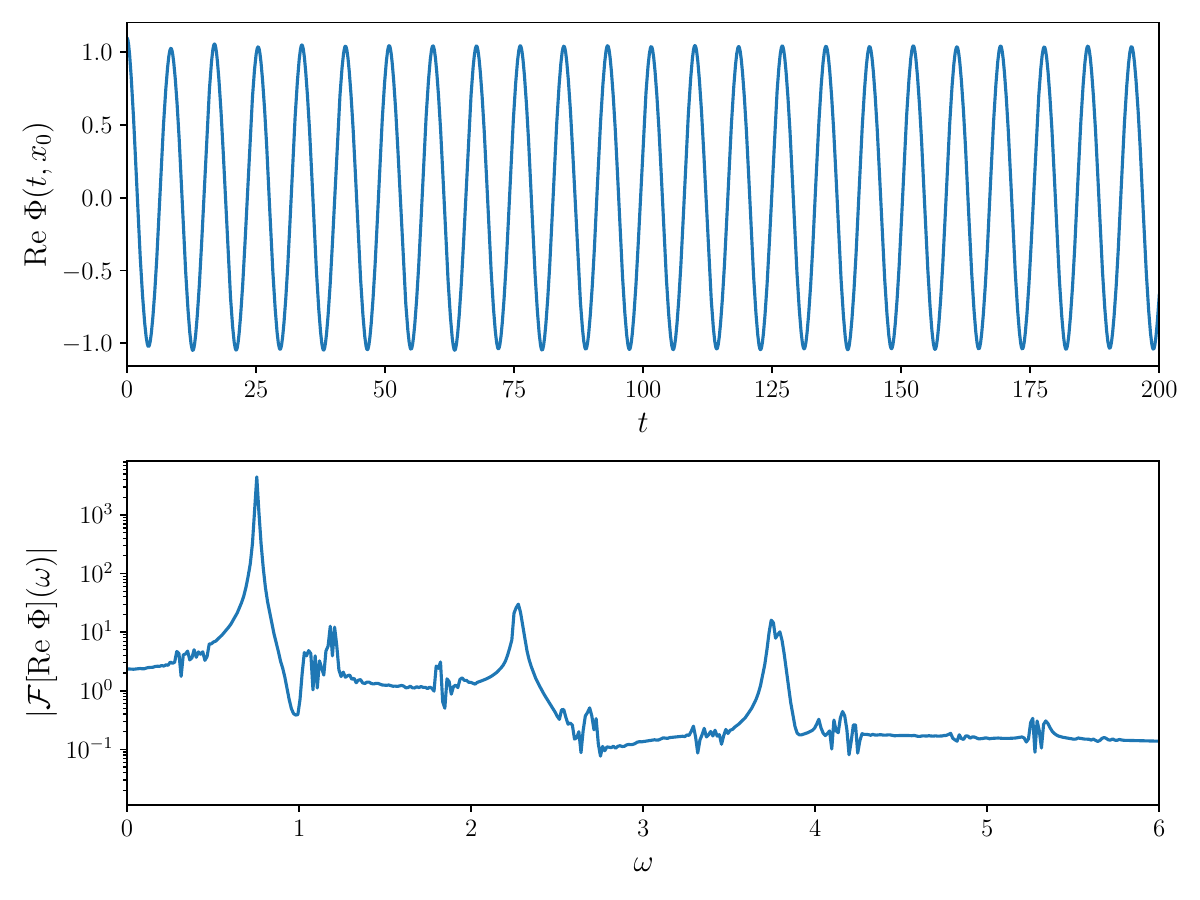}
    \includegraphics[width=0.48\linewidth]{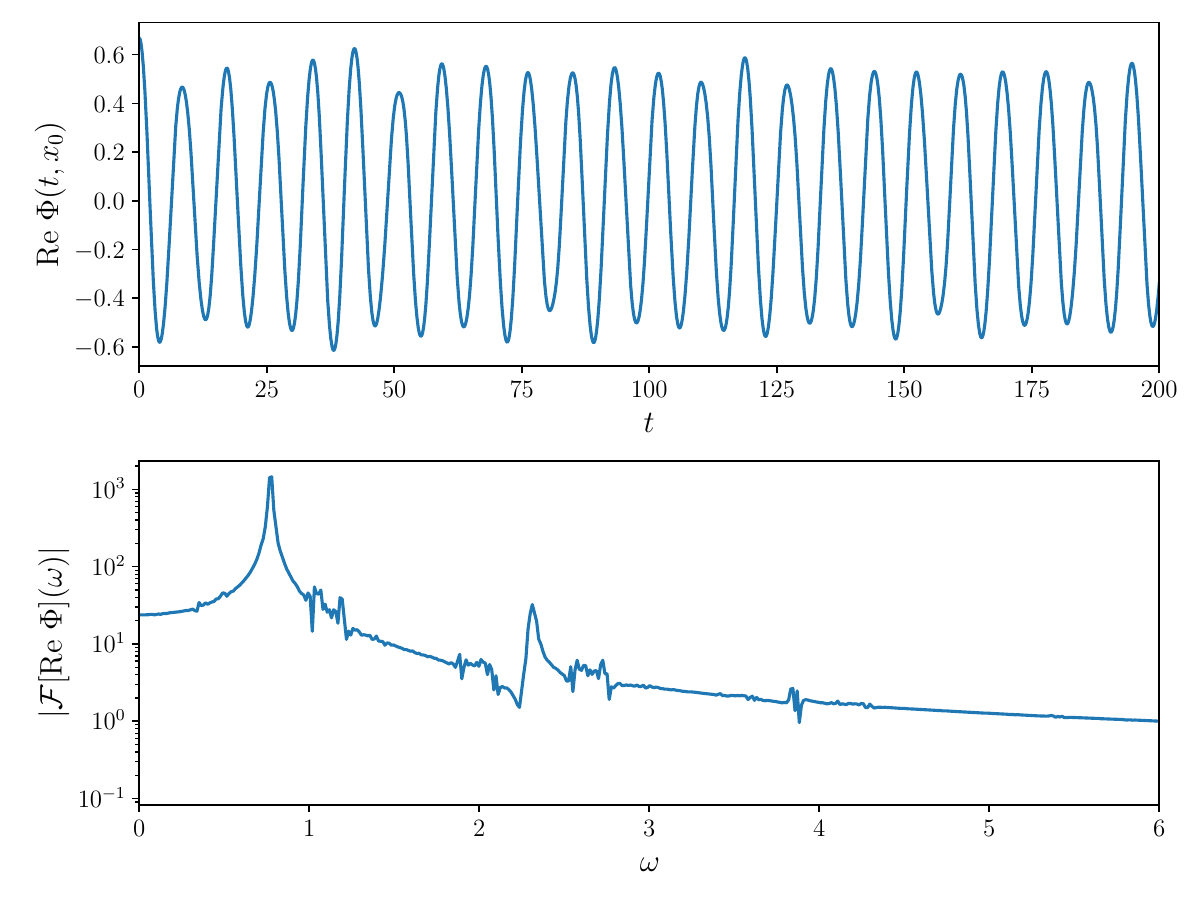}
    \includegraphics[width=0.48\linewidth]{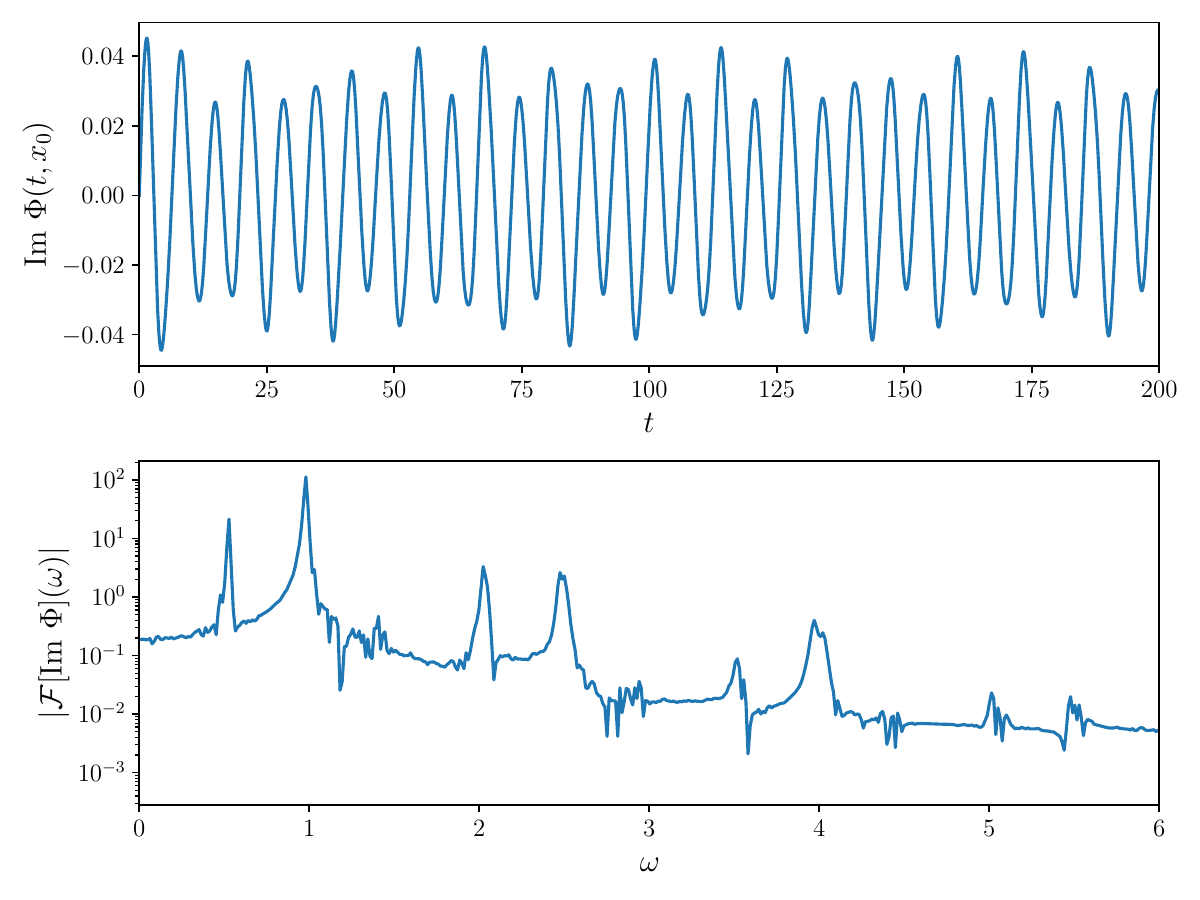}
    \includegraphics[width=0.48\linewidth]{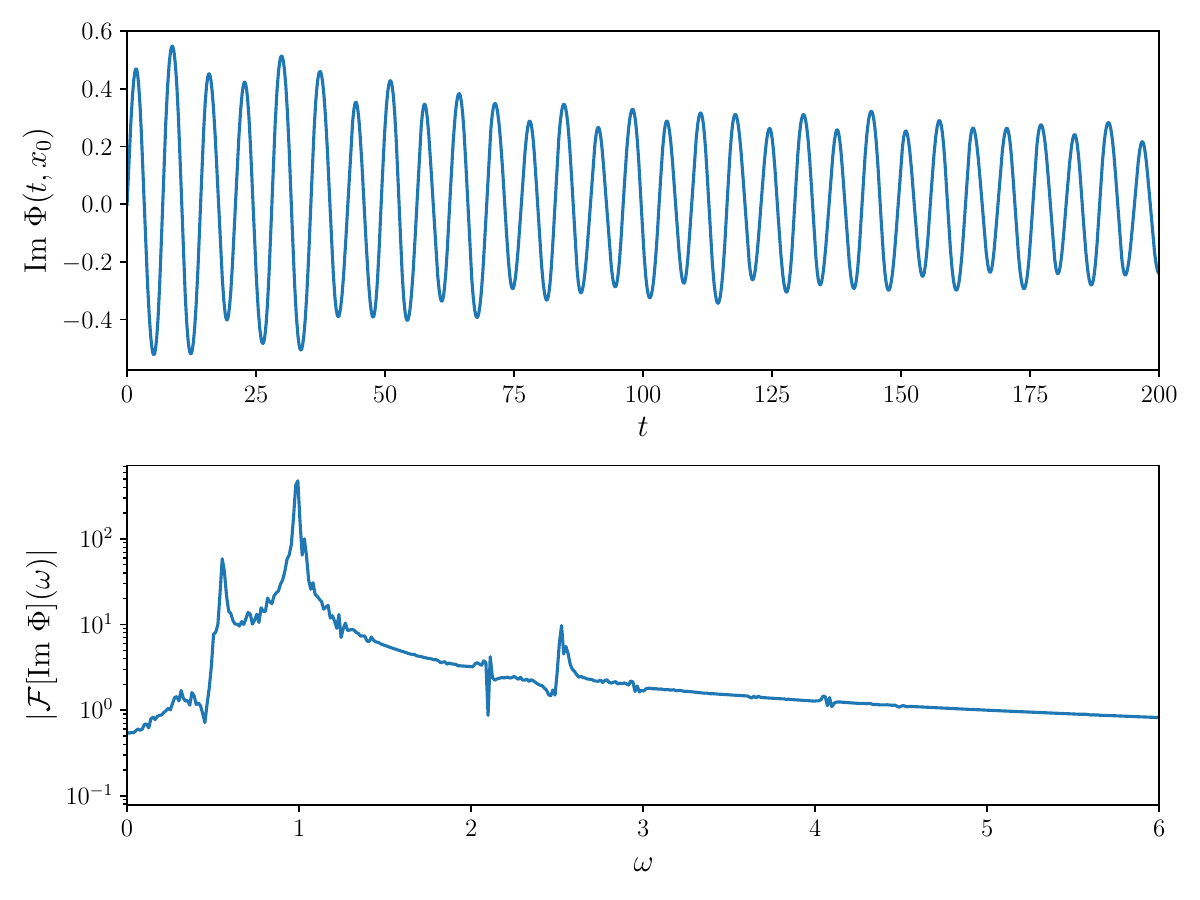}
    \caption{Example of the charge swapping solution obtained from the $QQ^*$ initial data (\ref{QQ*-ini}) with $x_0=0.5$ (left) and $x_0=2.5$ (right). Here $\beta=0.5$ and $\omega_0=0.85$. We plot charge density $\rho$, Re $\phi $ and Im $\phi$ at $x=x_0$ as well as the corresponding power spectra. }
    \label{fig:CS-example}
\end{figure}
It is important to note that although for $x_0\to 0$ we always see charge swapping, the amount of charge that actually is exchanged is decreasing to zero. Similarly, the imaginary component of the complex field is also less excited as the separation parameter in the initial configuration tends to 0. Eventually, in the limit $x_0\to 0$ we find an excited (modulated) oscillon which has zero charge density. Of course, this is a consequence of the form of the initial configuration. As $x_0$ approaches  zero it is just
\begin{equation}
    \Phi(x,t)=2f_{\omega_0}(x) \cos(\omega_0 t) + 2ix_0 f_{\omega_0}'(x) \sin (\omega_0 t). 
\end{equation}
Exactly for $x_0=0$ only the real component of the field is excited. Then, the initial configuration, $\Phi(x,t)=2f_{\omega_0} (x) \cos(\omega_0 t)$, relaxes to a real oscillon with an even perturbation. A small $x_0$ provides an additional odd perturbation of the oscillon along the imaginary component of the field.

In order to better understand the change-swapping solutions, we show the dependence of the power spectrum, computed for the solution generated from the initial data, on the separation parameter $x_0$, see Fig. \ref{fig:CS-scan}. This plot further confirms the existence of the three regimes. First of all, for large separation, $x_0 > 7.2$, the $Q$-ball do not have time to collide ($t_{max}=500$). Consequently, the power spectrum reveals the typical structure of an excited $Q$ ball, where the excitation comes from the presence of the second $Q$ ball. In fact, there are two main peaks: one at the internal frequency $\omega=\omega_0$ and another one corresponding to the quasinormal mode (red dashed line). For smaller separations, $7.2<x_0<2.9$, the $Q$-balls collide. The irregular multi-bounce regime is visible as a diffuse region in the power spectrum where the peak related to the internal frequency $\omega_0$ rapidly widens and gradually disappears from the spectrum.

Finally, for $x_0<2.9$, the initial configuration relaxes to the proper charge-swapping solution. Surprisingly, {\it the mode spectrum is not inherited from the $Q$-balls but from the excited oscillon}, realized at $x_0=0$. The spectrum is $\mathbb{Z}_2$ symmetric, a characteristic feature of an oscillon, with the leading peak at the fundamental oscillon frequency $\omega_O=0.7540$. This frequency changes very little over the whole charge swapping regime. There are additional peaks which also move slightly while we change $x_0$. The peaks can be easily explained by analyzing the power spectrum of the real and imaginary parts of the field separately, see Fig. \ref{fig:CS-example}. 


The fundamental frequency is visible as the main peak in the spectrum of the real component. This is due to the fact that at $x_0=0$ we find the real-valued oscillon. It changes from $\omega_{\rm Re}=0.7540$ at $x_0=0.5$ to $\omega_{\rm Re}=0.7791$ at $x_0=2.5$, which is close to the end of the charge-swapping regime. This frequency is absent in the spectrum of the imaginary part. Here, the main peak is located just below the mass threshold, that is, $\omega_{\rm Im}=0.9801$ for $x_0=0.5$ and $\omega_{\rm Im}=0.9927$. As originally observed in \cite{Xie:2021glp}, these two frequencies trigger charge swapping. Its frequency is simply $\omega_{CS}=\omega_{\rm Im}-\omega_O$ and is seen as the main peak in the power spectrum of the charge density at $\omega_{CS}=0.2262$ for $x_0=0.5$ and at $\omega_{CS}=0.2136$ for $x_0=2.5$. The second main peak corresponds to $\omega_O+\omega_{\rm Im}$.

\subsection{Charge-swapping from Generic Perturbed Oscillon Initial Conditions}

The small $x_0$ expansion also suggests that a charge-swapping solution can be generated if we start with a generic imaginary-valued perturbation of the real oscillon, not necessarily related to the split $QQ^*$ configuration. In fact, we found charge-swapping considering a general complex-valued perturbation of the oscillon in the real component of the complex field
\begin{equation}
    \phi(x,0) = \phi(x) + A_r g_1(x), \;\;\; \partial_t \phi(x,0)= i \omega A_i g_2(x), \label{init-gen}
\end{equation}
where $\phi(x)=\frac{2}{\sqrt{3}}\frac{\epsilon}{\cosh (\epsilon x)}$ is an approximated profile of the oscillon while the perturbation profiles are
\begin{equation}
    g_1(x)= \frac{1}{\cosh (\epsilon x/4)}, \;\;\; g_2(x)= \frac{\tanh(\epsilon x/4) }{\cosh (\epsilon x/4)}.
\end{equation}
$A_{r,i}$ are the amplitudes of the perturbations in the real and imaginary component.  We assume that the perturbation is wider than the original profile of the oscillon. 
\begin{figure}
    \centering
    \includegraphics[width=1.0\linewidth]{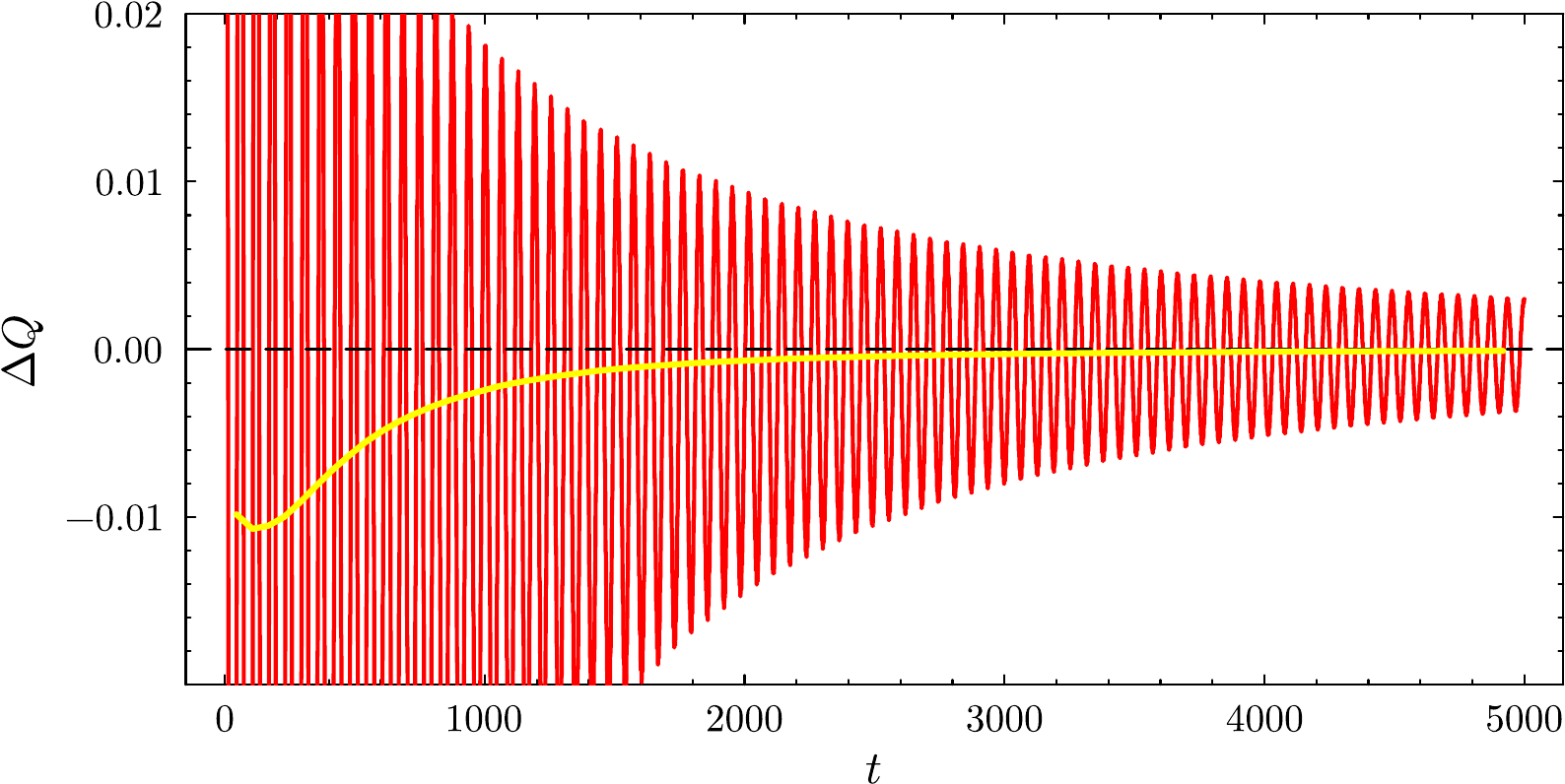}
    \caption{Time evolution of charge stored in the half-plane $x>0$ for the perturbed oscillon initial conditions (\ref{init-gen}) with $A_r=0$ and $A_i=0.1$. Yellow curve denotes the average charge around which the charge oscillates.}
    \label{fig:partail-CS}
\end{figure}

For $A_r\neq0$ and $A_i=0$ we obtain a standard perturbed real oscillon with some characteristic amplitude modulation. There is no excitation in the imaginary part of the complex field. Thus, the charge density vanishes identically.
\begin{figure}
    \centering
    \includegraphics[width=0.92\linewidth]{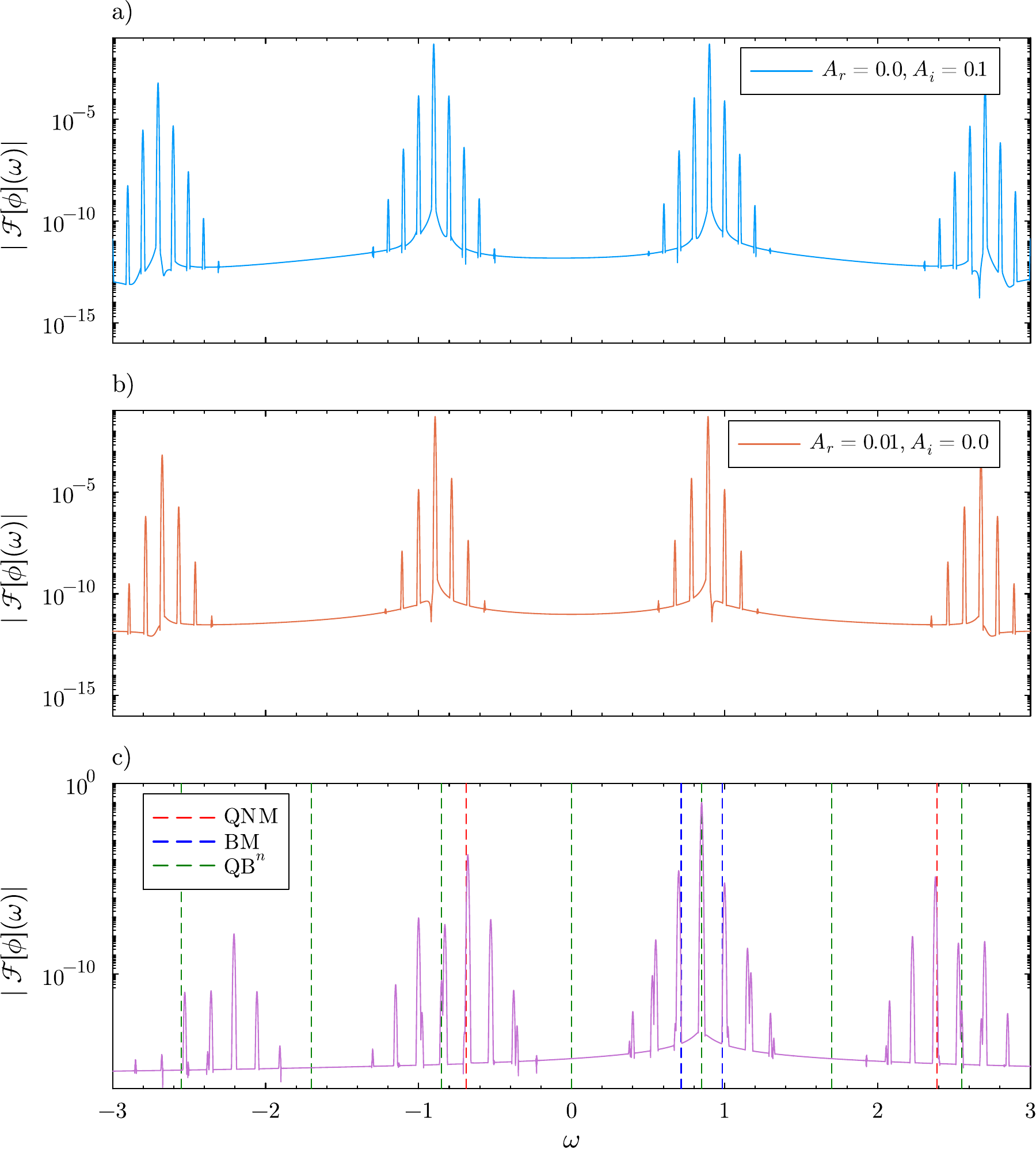}
    \caption{Power spectra of the the perturbed oscillon initial conditions (\ref{init-gen}) (upper and middle panels) compared with a perturbed $Q$-ball (bottom), where the quasi-normal mode (QNM), bound mode  (BM) and fundamental $Q$-ball frequency are marked.}
    \label{fig:partail-CS-modes}
\end{figure}

On the other hand, if we excite the imaginary component of the field, then charge density has a non-trivial distribution, even though the total charge is still zero. What is more important, these initial data amounts to the charge-swapping phenomenon. Again, the initial perturbation excites some oscillations in the imaginary part of the field with a frequency different from the frequency of the perturbed real-component oscillon. This difference leads to charge oscillations. 
However, generically, not the full charge is swapped between the $x>0$ and $x<0$ regions. This is clearly seen in Fig. \ref{fig:partail-CS} where we show the time dependence of $\Delta Q$. This charge oscillates around a negative value, which asymptotically tends to 0 recovering the usual full charge-swapping. We remark that one can fine-tune the initial data to obtain the full charge-swapping from the very beginning. 

In Fig. \ref{fig:partail-CS-modes} we present the power spectra of solutions obtained from the perturbed oscillon. As expected for the oscillon, the spectrum consists of a set of equally separated peaks, which reveals a $\mathbb{Z}_2$ symmetry. This should be contrasted with the spectrum of a perturbed $Q$-ball (squeezing by the factor $\lambda=1.05$), see the bottom panel. 

All this proves that the charge-swapping phenomenon is the result of a perturbation of the oscillon in the perpendicular direction (in the sense of the target space) rather than a $QQ^*$ bound state. 
\section{The collective coordinate model}
To obtain a collective description of the charge-swapping we need to have a set of configurations that occur in this phenomenon, at least approximately. Therefore, a natural choice would be to take an approximate oscillon perturbed in the imaginary direction, e.g. (\ref{init-gen}). From the above analysis, it is obvious that such configurations can also be achieved using a $QQ^*$ superposition with appropriately rescaled profiles of the $Q$-balls. This induces two collective coordinates: the positions of the $Q$-balls, $x_0$, and their amplitude $\alpha$. This has the advantage that it smoothly interpolates between a perturbed oscillon and a separated $QQ^*$ pair, and therefore it might even cover the formation of the charge-swapping solution from colliding $Q$-balls. In addition, in the single $Q$-ball sector, the collective variables $\alpha$ and $
\theta$ describe the excitation of the $Q$-ball. They must be included simultaneously due to charge conservation \cite{Battye:2000qj, Bowcock}.

All these motivate the following set of configurations 
\begin{equation}
    \Phi(x)=\alpha \left(f_\omega(x+x_0) e^{i\theta} + f_\omega(x-x_0) e^{-i\theta} \right), \label{set}
\end{equation}
where the collective coordinates (moduli) $(X^1,X^2,X^3)=(\alpha,\theta, x_0)$ will be promoted to time-dependent variables \cite{Manton:1981mp}. Furthermore, $f_\omega$ is the profile of the $Q$-ball with frequency $\omega$. This choice defines the initial condition for the velocity of the coordinate $\theta$. In the CCM framework, the full dynamics will be approximated by the dynamics of the moduli. In general, this set leads to a three-dimensional collective coordinate model (CCM).

In the current paper, we will consider its two-dimensional version, where only $(X^1,X^2)=(\alpha,\theta)$ are dynamical variables, while $X^3=x_0$ is treated as a fixed parameter. This greatly simplifies computations and still gives very accurate results. Thus, as the separation is kept constant, this version cannot describe the coalescence of the $Q$-ball with the anti-$Q$-ball. However, it is able to cover dynamics in each sector separately and predict at which separation the charge-swapping state shows up. 

The Lagrangian of this CCM reads
\begin{equation}
    L[\alpha, \theta] = g_{ij} \dot{X}^i \dot{X}^j - V(X^1,X^2),
    \label{moduli}
\end{equation}
where the moduli space metric is
\begin{equation}
    g_{ij} = \left(
\begin{array}{cc}
     \frac{Q}{\omega} +2\mathcal{I} \cos(2\theta)  & - 2 \mathcal{I} \alpha\sin(2\theta)   \\
   -2 \mathcal{I} \alpha \sin(2\theta)   & \alpha^2 \left(\frac{Q}{\omega} -2\mathcal{I} \cos(2\theta) \right)
\end{array}
    \right).
\end{equation}
Here 
\begin{equation}
    \mathcal{I}(x_0)=\int_{-\infty}^\infty f_\omega(x+x_0)f_\omega(x-x_0) dx
\end{equation}
measures the spatial overlap of the $Q$-ball and anti-$Q$-ball profiles. It is easy to show that $0\leq \mathcal{I} \leq \frac{Q}{2\omega}$. For infinitely separated solitons, $x_0\to \infty$, the overlap tends to 0. For the on-top configuration, that is, when $x_0\to 0$, we get $\mathcal{I}=\frac{Q}{2\omega}$. 

The effective potential $V$ is a sixth order polynomial in $\alpha$
\begin{equation}
    V(\alpha, \theta)=\sum_{j=0}^3 \sum_{k=1}^3 C_{k}^{(j)} \alpha^{2k} \cos (2j\theta),
\end{equation}
where $C_{k}^{(j)}$ are constants depending on $x_0$, which have to be computed numerically, see Appendix \ref{Eff_pot}. 

The CCM gives a well-defined two-dimensional dynamical system for any $x_0>0$. However, exactly at $x_0=0$, where the positions of the $Q$ balls coincide, we encounter a singularity. This can be seen in the determinant of the metric
\begin{equation}
    \mbox{det} g = \alpha^2 \left( \frac{Q^2}{\omega^2} - 4\mathcal{I}^2 \right)
\end{equation}
which is identically zero for $x_0=0$. The reason for this is that the set of complex configurations (\ref{set}) reduces to real functions at this point, effectively lowering the dimensionality of the moduli space for $x_0=0$. A consequence of this is an observation that the description of the oscillon within this CCM is highly simplified. Exactly at $x_0=0$, the CCM has only one dynamical coordinate. Thus, the oscillon is modeled in terms of a one-dimensional anharmonic oscillator. This obviously neglects several important features of the oscillons, e.g. modulation of the amplitudes, which requires at least two degrees of freedom \cite{Blaschke:2024dlt, Blaschke:2025qkg}, see also \cite{Manton:2023mdr, Navarro-Obregon:2023hqe}. Nevertheless, for our purposes, it is enough to include the oscillon in such a simplified way. Note that there is a second point at which the determinant of the metric vanishes, $\alpha=0$. However, it is an apparent coordinate system singularity, identical to the singularity of the polar coordinates.

There exists another convenient set of collective coordinates that removes the apparent singularity. Namely, $(X^1,X^2)=(X,Y)$ being the Cartesian coordinates on the complex target space
\begin{equation}
    X=\alpha \cos \theta, \;\;\; Y=\alpha \sin \theta. 
\end{equation}
Then, the real and imaginary parts of the complex field are simply
\begin{align}
    \mbox{Re} \, \Phi (x,t) &= \left(f_\omega(x+x_0)+f_\omega(x-x_0)\right) X (t), \nonumber 
 \\ \mbox{Im} \, \Phi (x,t) &= \left(f_\omega(x+x_0)-f_\omega(x-x_0)\right) Y (t).
\end{align}
Importantly, the corresponding metric is diagonal and constant
\begin{equation}
    g_{ij} = \left(
\begin{array}{cc}
     \frac{Q}{\omega} +2\mathcal{I} & 0  \\
   0  &  \frac{Q}{\omega} -2\mathcal{I} 
\end{array}
    \right).
\end{equation}
\noindent This smooth change of variables cannot affect the existence of the singularity at $x_0=0$. In fact, in this limit the $g_{YY}$ component of the metric is zero, which obviously leads to zero of the determinant. 

Since the metric is flat and diagonal, all the dynamics are encoded into the nonlinearities of the effective potential, which, written in the Cartesian coordinates, is an even polynomial of the sixth order and reads
\begin{figure}
    \centering
    \includegraphics[width=0.45\linewidth]{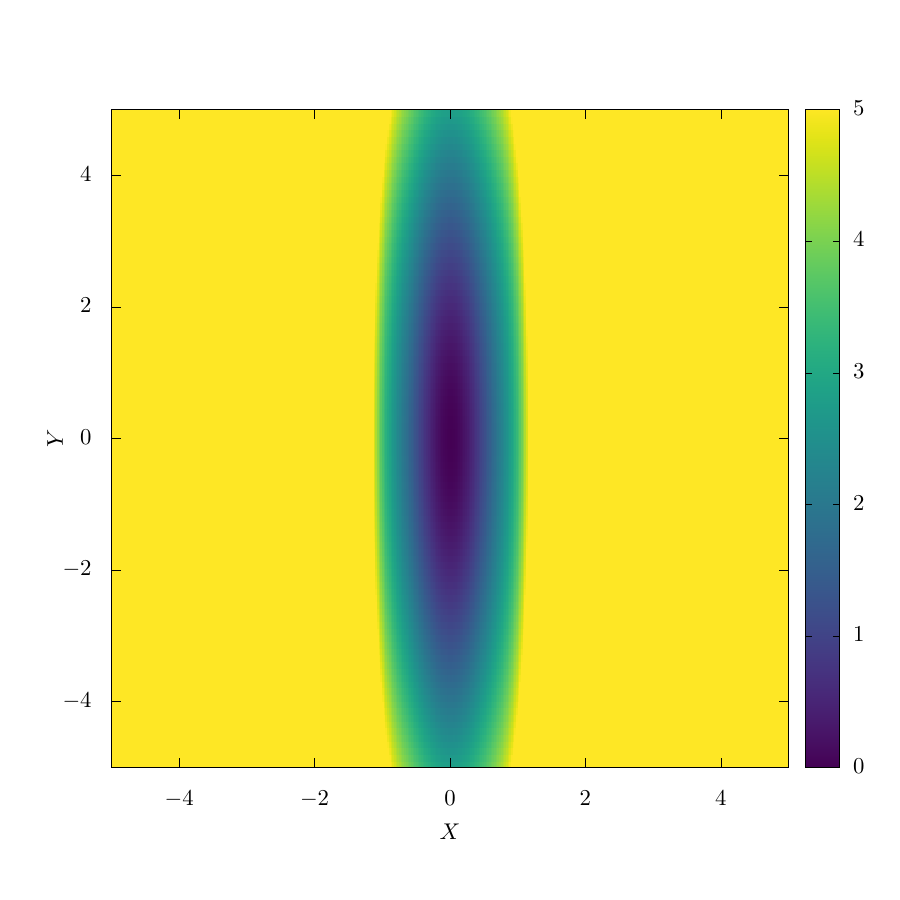}
    \includegraphics[width=0.45\linewidth]{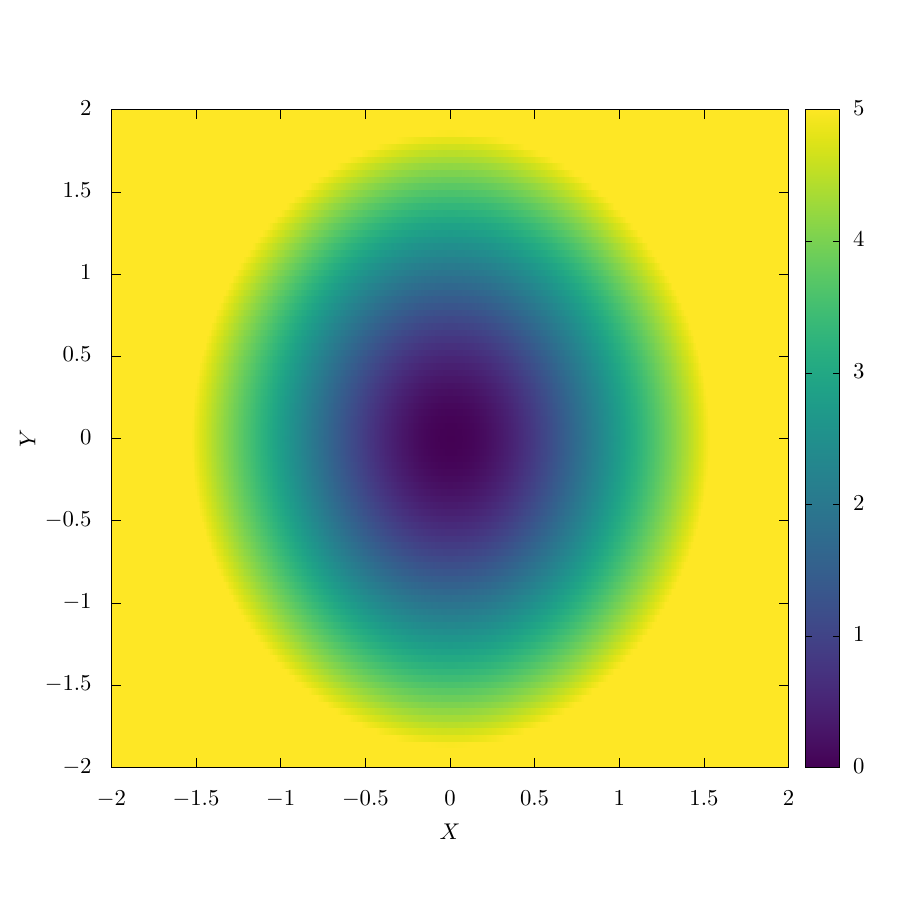}
    \caption{The effective potential V(X,Y). Left: $x_0=0.5$ i.e., the charge-swapping regime. Right: $x_0=3$ i.e., the two separated $Q$-balls regime. }
    \label{fig:eff_pot}
\end{figure}
\begin{align}
     V(X,Y)&= D_{2,0}X^2+D_{4,0}X^4+D_{6,0}X^6+
    D_{0,2}Y^2+D_{0,4}Y^4+D_{0,6}Y^6 \nonumber \\
    &+D_{2,2}X^2Y^2+D_{4,2}X^4Y^2+D_{2,4}X^2Y^4.
\end{align}
In Appendix \ref{Eff_pot}, we give integral expressions for these coefficients and their relation to the constants $C^{(j)}_k$. Qualitatively, the potential is simply a nonaxially symmetric anharmonic potential with only one minimum at $X=Y=0$. In the limit of separated $Q$ balls, when $x_0\to \infty$, it recovers the axial symmetry as both $Q$ balls independently rotate in the target space, although already for $x_0 \approx 3$ it has a quite symmetric shape, see Fig. \ref{fig:eff_pot}. In the limit where $x_0\to 0$ the potential becomes more and more elongate along the $Y$ direction. However, because the metric component $g_{YY}$ also vanishes in this limit, all excitations in this direction are suppressed. 

\begin{figure}
    \centering
    \includegraphics[width=0.32\linewidth]{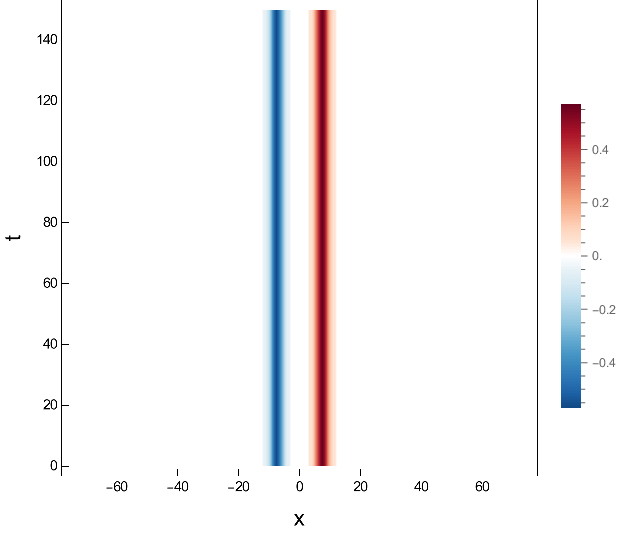}
    \includegraphics[width=0.32\linewidth]{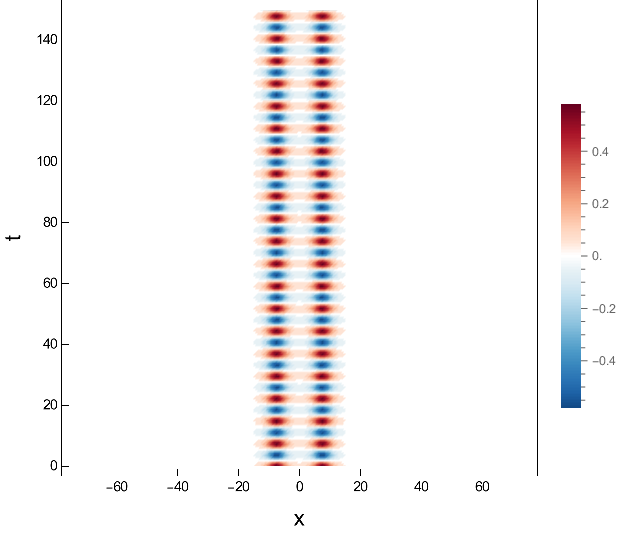}
    \includegraphics[width=0.32\linewidth]{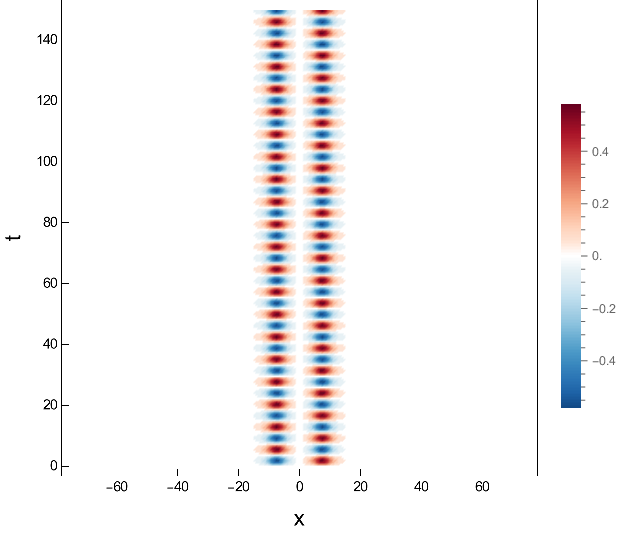}
    \includegraphics[width=0.32\linewidth]{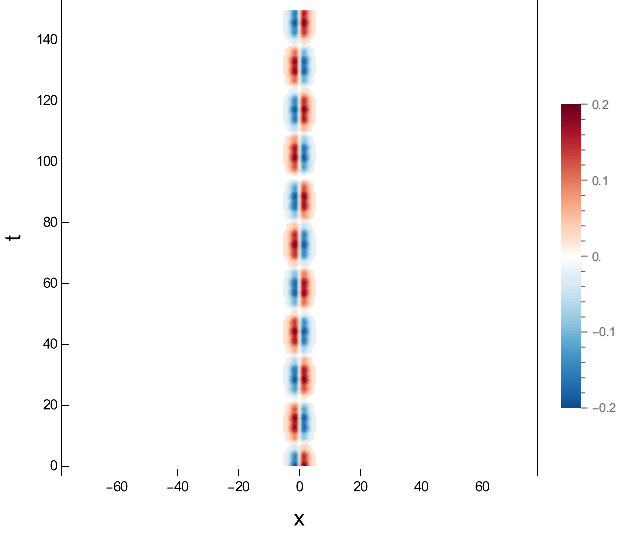}
    \includegraphics[width=0.32\linewidth]{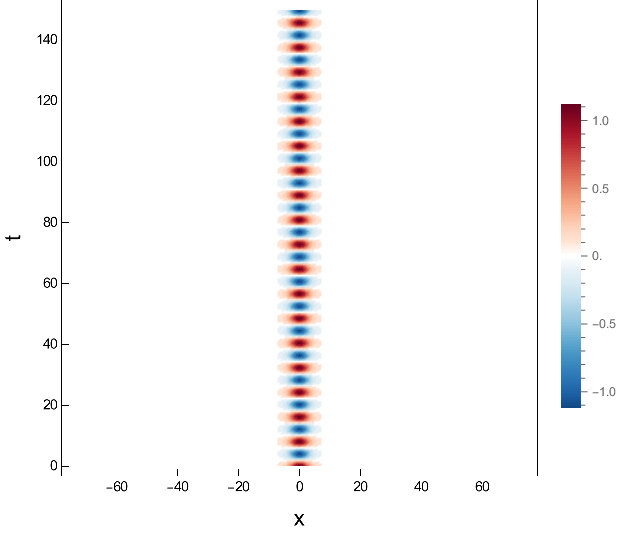}
    \includegraphics[width=0.32\linewidth]{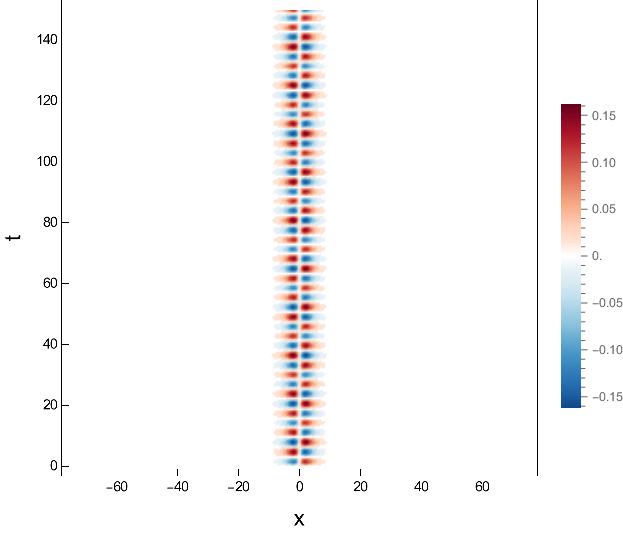}
    \caption{Examples of time evolution obtained from the CCM with $\beta=0.5$ and $\omega_0=0.85$. Here $x_0=7.5$ (upper row), and $x_0=0.5$ (lower row). We plot: the charge density (left column), Re $\phi$ (middle column) and Im $\phi$ (right column). }
    \label{fig:CS-exampl}
\end{figure}

To reproduce the full field theory dynamics we apply the CCM with the following initial conditions
\begin{align}
\alpha(0)&=1, \;\;\; \dot{\alpha}(0)=0  \\
\theta(0)&=0, \;\;\; \dot{\theta}(0)=\omega, 
\end{align}
that is 
\begin{align}
X(0)&=1, \;\;\; \dot{X}(0)=0  \\
Y(0)&=0, \;\;\; \dot{Y}(0)=\omega, 
\end{align}
which coincide with the choice of the set of configuration (\ref{set}).
\begin{figure}
    \centering
    \includegraphics[width=0.85\linewidth]{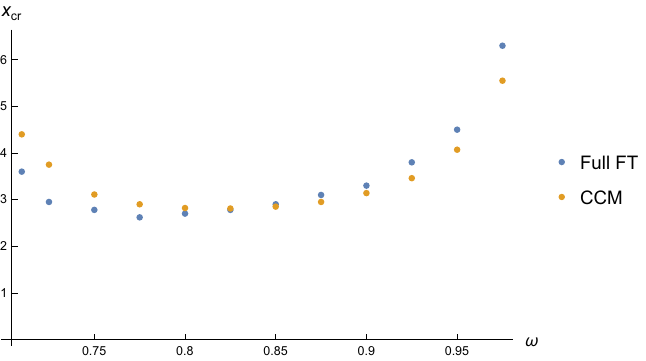}
    \caption{The dependence of the critical value of the parameter $x_{cr}$ on the frequency $\omega$ of the initial $QQ^*$ configuration. For $x_0<x_{cr}$ we find the charge-swapping solution. }
    \label{fig:border}
\end{figure}

Qualitatively, the CCM provides two types of solutions, see \ref{fig:CS-exampl}. For small $x_0$ we find charge swapping solutions while for bigger $x_0$ the $Q$-balls became basically independent. This, of course, reproduces the main property of full-field-theory dynamics. A consequence of this is the existence of a critical position (half of the separation) $x_{cr}$ that divides these two regimes. In Fig. \ref{fig:border} we compare the full field theory result with the CCM computation. Here $\beta=0.5$ and we vary the initial frequency $\omega$. The agreement is very good. Undoubtedly, such a simple CCM predicts the dependence of the critical distance $x_{cr}$ on $\omega$. 

\begin{figure}
    \centering
    \includegraphics[width=0.45\linewidth]{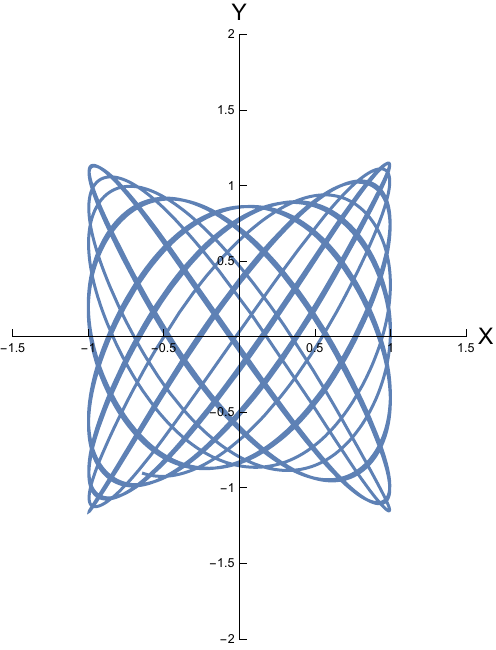} \hspace{1.5pt}
    \includegraphics[width=0.45\linewidth]{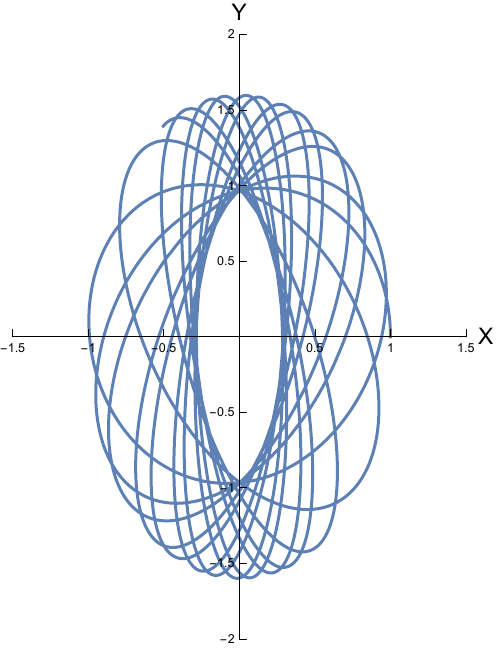}
    \caption{Phase portraits for $x_0=0.5$ in the CSQ stage (left) and $x_0=3$ in the two separated $Q$-balls stage (right). }
    \label{fig:eff_orbit}
\end{figure}

The existence of these two regimes is strongly related to the form of the effective potential $V$. If $x_0$ is large, the potential has an approximately axially symmetric shape, which amounts to the existence of approximately circular orbits in the $(X,Y)$ plane. Strictly speaking, the orbits are bounded inside to ellipses, see Fig. \ref{fig:eff_orbit}. Obviously, in the limit $x_0 \to \infty$, the ellipses tend to the same circle, and we find two (infinitely separated) $Q$-balls. For small $x_0$, where the charge-swapping solution is reproduced, the shape of the potential is strongly elongated and instead of quasi-circular orbits we find orbits inside a rather rectangular region. This was previously observed in \cite{Xie:2021glp}. Interestingly, the critical separation has a geometric interpretation. For $x_0=x_{cr}$, the phase portrait forms a "hole" at the origin, which indicates a transition from the rectangular to two-ellipse case. 

\begin{figure}
    \centering
     \includegraphics[width=0.48\linewidth]{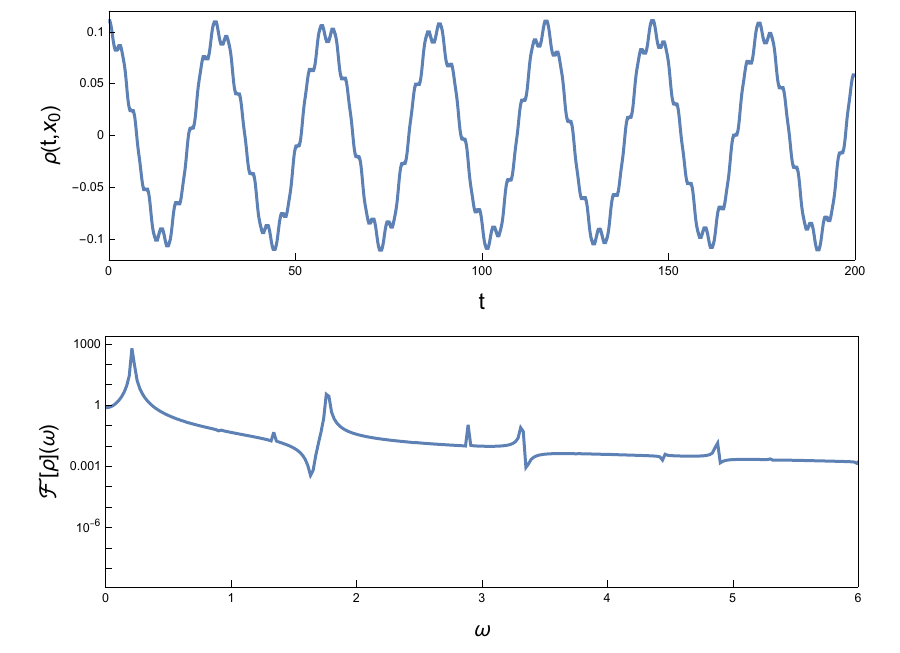}
     \includegraphics[width=0.48\linewidth]{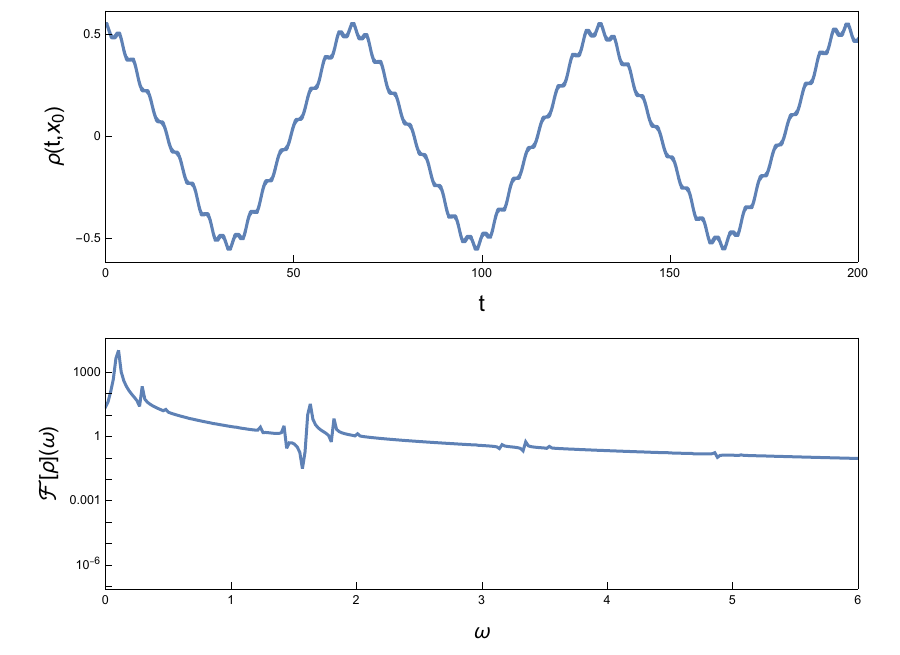}
     
     \includegraphics[width=0.48\linewidth]{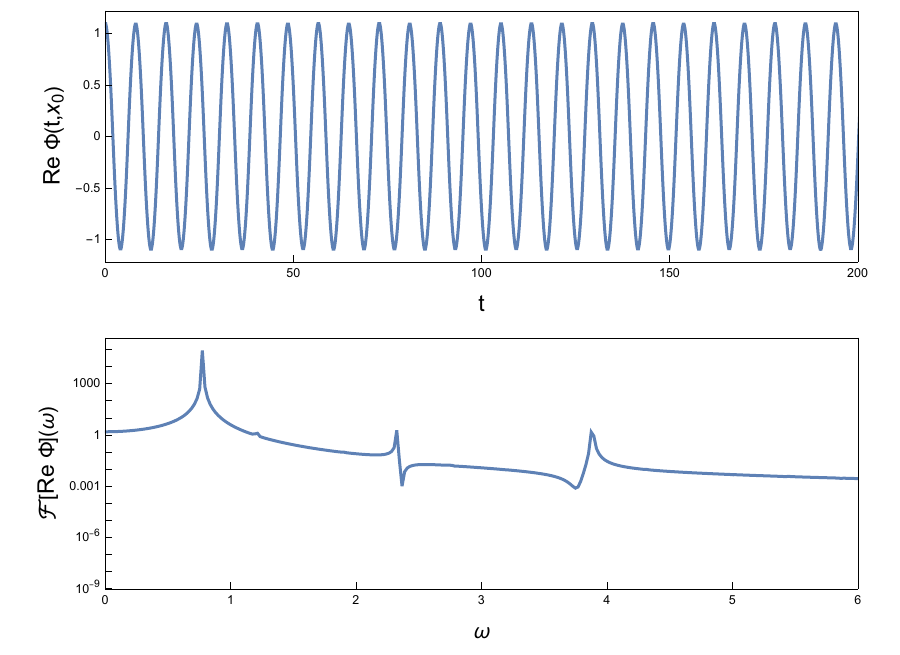}
     \includegraphics[width=0.48\linewidth]{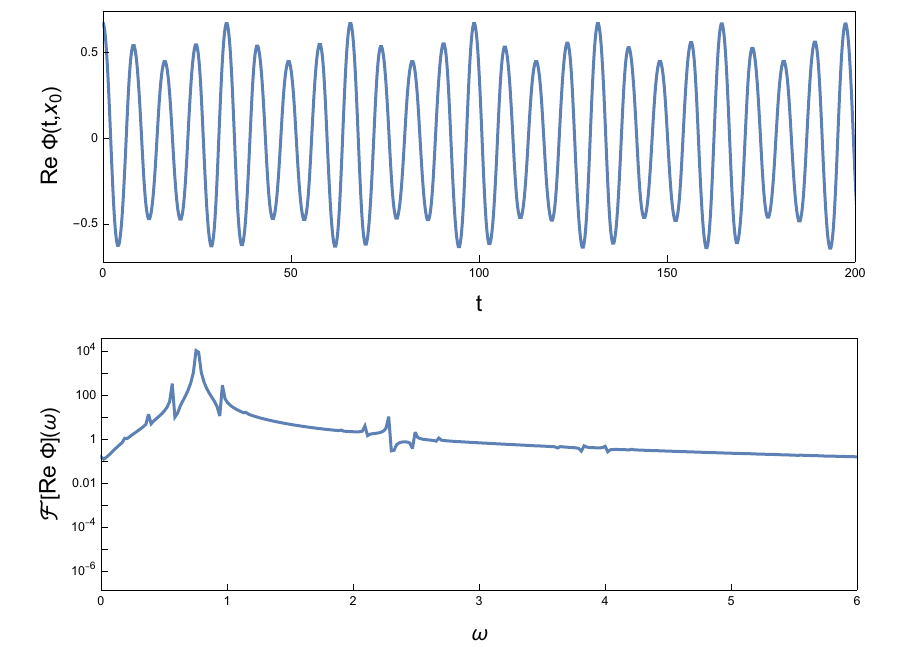}
    
     \includegraphics[width=0.48\linewidth]{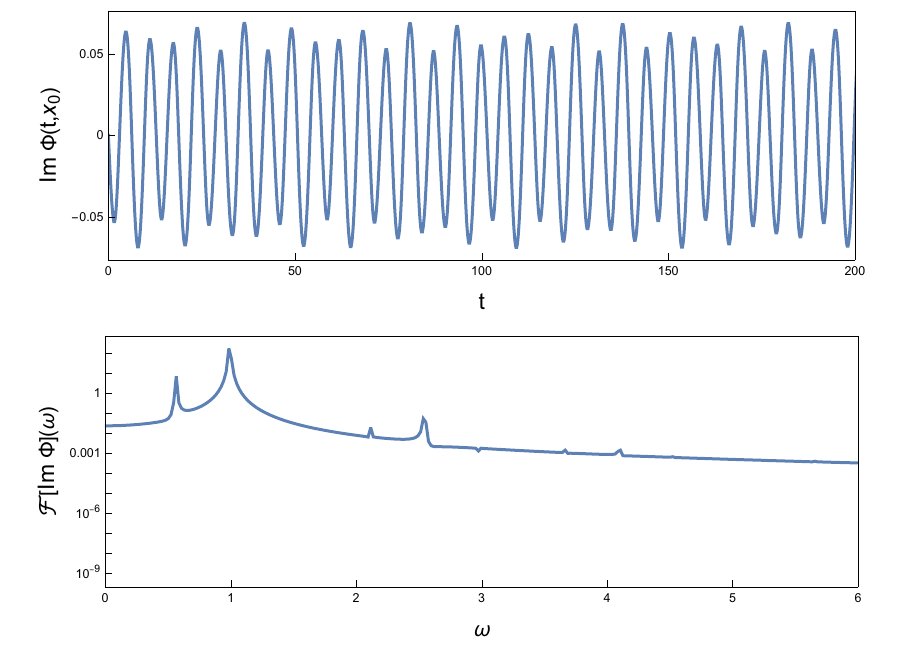}
     \includegraphics[width=0.48\linewidth]{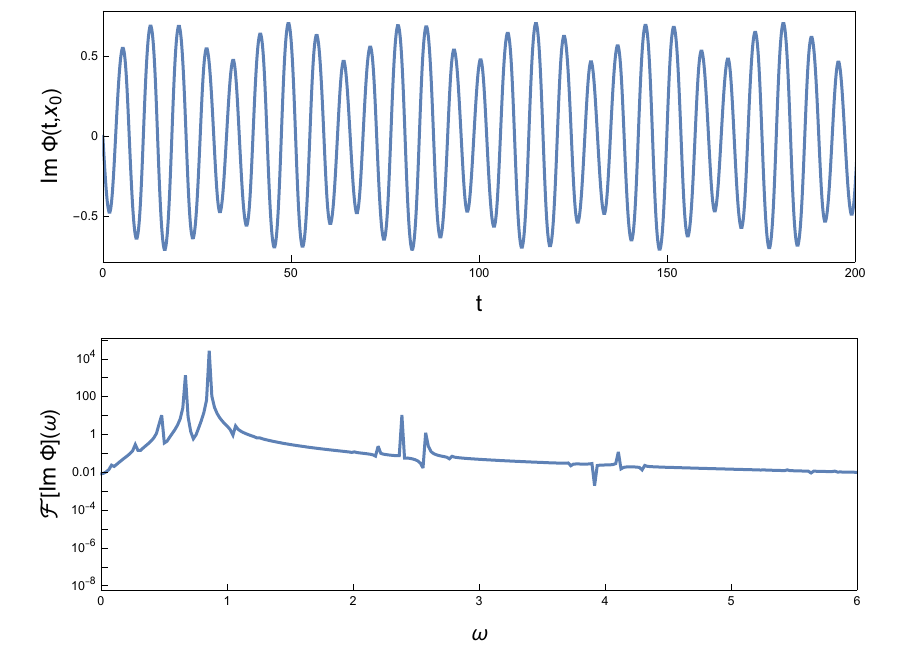}
    \caption{Charge-swapping solutions obtained from the CCM with $x_0=0.5$ (left column) and $x_0=2.5$ (right column). Here $\beta=0.5$ and $\omega_0=0.85$. We plot charge density $\rho$, Re $\phi $ and Im $\phi$ at $x=x_0$ as well as the corresponding power spectra. }
    \label{fig:CCM-example}
\end{figure}

In Fig. \ref{fig:CCM-example} we present the dynamics obtained in the CCM (\ref{moduli}) for $\beta=0.5$ and $\omega_0=0.85$. Specifically, we chose $x_0=0.5$ and $x_0=2.5$ which corresponds to the initial data for the full field theory numerics shown in Fig. \ref{fig:CS-example}. We found good agreement, especially for the case with small $x_0$. The amplitudes of oscillations of the real and imaginary components of the field agree well with the full theory numerics although, generically, they are bigger. This is best visible in the amplitude of the charge-swapping. This is due to the fact that the initial configuration radiates quite a lot at the initial stage of the full field-theoretical evolution. This is, of course, not possible in the CCM since no radiative degrees of freedom are included. Hence, all the initial energy goes to the oscillating solutions.

Although the CCM predicts amplitudes of the field and charge density that are larger than those observed, it accurately reproduces the frequencies. That is, the fundamental frequency of the perturbed oscillon is $\omega_{\rm Re}=0.7749$, for $x_0=0.5$ and $\omega_{\rm Re}=0.7540$ for $x_0=2.5$, which are very close to the values found in the full field theory. The main frequency of the imaginary component is more sensitive to the value of $x_0$. For $x_0=0.5$, we get $\omega_{\rm Im} = 0.9840$, which is quite close to the true value. However, for $x_0=2.5$, the discrepancy is much larger, $\omega_{\rm Im} = 0.8587$. In any case, this leads to a good prediction of the charge-swapping frequency, especially in the small $x_0$ limit -  $\omega_{CS}=0.2095$ for $x_0=0.5$. For larger $x_0$, disagreement grows; for example, for $x_0=2.5$, we find $\omega_{CS}=0.1047$. 

\section{Summary}

Since its discovery, the charge-swapping phenomenon has been treated as a bound state of $Q$-ball and anti-$Q$-ball. Despite the fact that it arises from such an initial state, we have clearly shown that it should rather be understood as an excited oscillon; that is, an oscillon with a charged wave. The main argument supporting this conclusion comes from the analysis of the mode structure of the charge-swapping solutions, which is basically the same as the spectrum of the oscillon. In addition, the charge-swapping solution is easily formed from perturbed oscillon initial data, where the perturbation is along the imaginary direction. 

In addition, we constructed a collective coordinate model (CCM) that captures the essence of the charge-swapping mechanism. Even in the simplest two-dimensional version, it is able to model the charge-swapping solutions with reasonably good accuracy. It predicts the frequencies of the field oscillations as well as the charge-swapping frequency, especially well in the limit of small imaginary perturbation of the real oscillon, that is, in the small $x_0$ limit. We expect that the three-dimensional CCM, where the parameter $x_0$ becomes a dynamical variable, will significantly improve the collective description. However, this requires a careful treatment of the singularity in $x_0=0$. A natural approach to this issue can be based on the perturbative relativistic moduli space framework developed in \cite{Adam:2021gat} and successfully applied to kink-antikink collisions. 

    In our analysis, we also found evidence that the interaction of the $Q$-balls reveals a chaotic pattern, very similar to collisions of domain walls (kinks) \cite{Campbell:1983xu, Sugiyama:1979mi, Manton:2021ipk} and local vortices \cite{Krusch:2024vuy, AlonsoIzquierdo:2024nbn}. Similarly as in the preceding cases, this behaviour is probably triggered by the resonant energy transfer between the kinetic and internal degrees of freedom, that is, the bound modes or quasi-normal modes of the $Q$-ball. As we saw in Fig. \ref{fig:CS-scan}, the Feshbach resonance is especially easily excited if the $Q$-ball is perturbed by the presence of another $Q$-ball. Hence, it may participate in the resonant energy transfer mechanism. In fact, the Feshbach resonances are known to play a significant role in kink dynamics \cite{GarciaMartin-Caro:2025zkc}. This should be further studied. 

Looking from a wider perspective, the oscillon origin of the charge-swapping once again underlines the very close relation between oscillons and $Q$-balls in complex field theories. In fact, as recently discovered, these objects are connected by an approximate duality \cite{Blaschke:2025anm}. Speaking precisely, an oscillon can be viewed as a bound state of $Q$-ball and anti-$Q$-ball, located on top of each opther. Consequently, the charge-swapping phenomenon may also find a dual description as a four-Q-ball state.

\section*{Acknowledgements}
The authors acknowledge support from the Spanish Ministerio de Ciencia e Innovacion (MCIN) with funding from the European Union NextGenerationEU (Grant No. PRTRC17.I1) and the Consejeria de Educacion from JCyL through the QCAYLE project, as well as the grant
PID2023-148409NB-I00 MTM.
K. S. acknowledges
financial support from the Polish National Science
Centre (Grant No. NCN 2021/43/D/ST2/01122). We thank Paul Saffin for comments. 

\appendix
\section{Effective potential} \label{Eff_pot}
The coefficients in the potential of the CCM are related as follows
\begin{align}
     D_{2,0}&=C^{(0)}_{1}+C^{(1)}_{1}, \;\;\; D_{0,2}=C^{(0)}_{1}-C^{(1)}_{1}, \\
     D_{4,0}&=C^{(0)}_{2}+C^{(1)}_{2}+C^{(2)}_{2}, \;\;\; 
     D_{0,4}=C^{(0)}_{2}-C^{(1)}_{2}+C^{(2)}_{2},\;\;\;  D_{2,2}=2C^{(0)}_{2}-6C^{(2)}_{2}, \\
      D_{6,0}&=C^{(0)}_{3}+C^{(1)}_{3}+C^{(2)}_{3}+C^{(3)}_{3}, \;\;\;
      D_{0,6}=C^{(0)}_{3}-C^{(1)}_{3}+C^{(2)}_{3}-C^{(3)}_{3}, \\
      D_{4,2}&=3C^{(0)}_{3}+C^{(1)}_{3}-5C^{(2)}_{3}-15C^{(3)}_{3},
      D_{2,4}=3C^{(0)}_{3}-C^{(1)}_{3}-5C^{(2)}_{3}+15C^{(3)}_{3},
\end{align}
where the non-zero constants $C^{(j)}_k$ are given by the following formulas 
\begin{align}
        C^{(0)}_1 &
        = \int dx \left( f_\omega^\prime(x+x_0)^2 + f_\omega(x+x_0)^2 \right) + \int dx \left(  f_\omega^\prime(x-x_0)^2+ f_\omega(x-x_0)^2 \right),\\
        C^{(0)}_2 & = -\int dx \left( f_\omega(x+x_0)^4 +f_\omega(x+x_0)^2f_\omega(x-x_0)^2 + f_\omega(x-x_0)^4\right), \\
        C^{(0)}_3 & = \beta\int dx \left( f_\omega(x+x_0)^6 + 9 f_\omega(x+x_0)^4f_\omega(x-x_0)^2 \right. \\
        & \left. \hspace*{2.0cm} + 9 f_\omega(x+x_0)^2f_\omega(x-x_0)^4 + f_\omega(x-x_0)^6\right),  \\
        C^{(1)}_1 & = \int dx \left( f_\omega^\prime(x+x_0)f_\omega^\prime(x-x_0) + f_\omega(x+x_0)f_\omega(x-x_0)\right),\\
        C^{(1)}_2 & = -4   \int dx \left( f_\omega(x+x_0)^3 f_\omega(x-x_0) +   f_\omega(x+x_0)   f_\omega(x-x_0)^3 \right),\\
        C^{(1)}_3 &  = 6\beta \int dx \left( f_\omega(x+x_0)^5 f_\omega(x-x_0) \right. \\
        &\left. \hspace*{2.0cm} + 3 f_\omega(x+x_0)^3 f_\omega(x-x_0)^3 + f_\omega(x+x_0) f_\omega(x-x_0)^5\right),\\
       C^{(2)}_2 & = -2   \int dx \left( f_\omega(x+x_0)^2f_\omega(x-x_0)^2 \right), \\
        C^{(2)}_3 & = 6\beta \int dx \left( f_\omega(x+x_0)^4f_\omega(x-x_0)^2 + f_\omega(x+x_0)^2f_\omega(x-x_0)^4 \right),\\
         C^{(3)}_3 & = 2\beta \int dx \left( f_\omega(x+x_0)^3 f_\omega(x-x_0)^3 \right).\\
\end{align}
In the limit $x_0 \rightarrow \infty$, the potential $V(\alpha, \theta)$ becomes independent of $\theta$, as only the terms proportional to $C_k^{(0)}$ remain non-zero:
\begin{equation}
    V(\alpha) = \sum_{k=1}^{3}C_k^{(0)}\alpha^{2k}=C_1^{(0)}\alpha^2+C_2^{(0)}\alpha^4+C_3^{(0)}\alpha^6.
\end{equation}
Therefore, the potential becomes axially symmetric in this limit.
\begin{figure}
    \centering
    \includegraphics[width=0.9\linewidth]{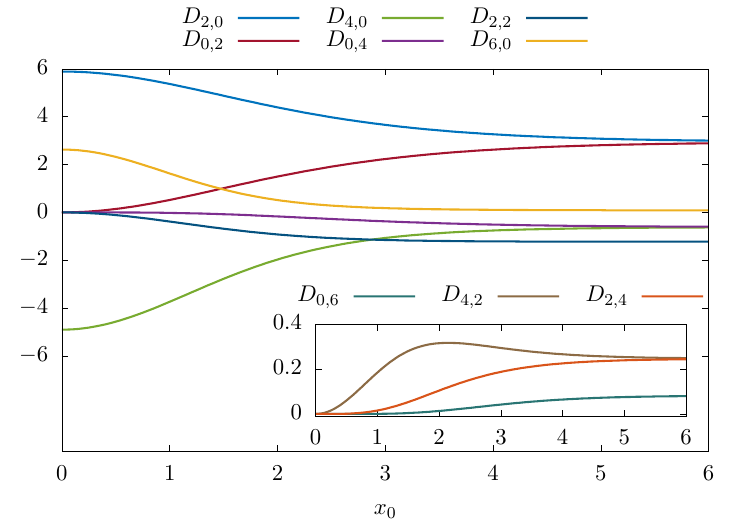}
    \caption{Dependence of the couplings in the effective potential on the position of the centres $x_0$.}
    \label{fig:potential_D}
\end{figure}

By making the change to Cartesian coordinates, the new coefficients $D$ can be written in terms of $C^{(0)}_k$:
\begin{align}
   & D_{2,0} = D_{0,2} = C_1^{(0)} = 2\int dx \left( f_\omega'(x)^2+f_\omega(x)^2\right) \ , \\
   & D_{4,0} = D_{0,4} = \dfrac{1}{2} D_{2,2} = C_2^{(0)} = -2\int dx \ f_\omega(x)^4 \ , \\
   & D_{6,0} = D_{0,6} = \dfrac{1}{3} D_{4,2} = \dfrac{1}{3} D_{2,4} = C_3^{(0)} = 2 \beta \int dx \ f_\omega(x)^6.
\end{align}

In Fig. \ref{fig:potential_D} we plot the dependence of the coefficients $D$ on $x_0$. The asymptotic values of the parameters $D$ at $x_0\to \infty$ are
\begin{align}
   & D_{2,0} = D_{0,2} = C_1^{(0)} = 2.94398 \\
   & D_{4,0} = D_{0,4} = \dfrac{1}{2} D_{2,2} = C_2^{(0)} = -0.612644 \\
   & D_{6,0} = D_{0,6} = \dfrac{1}{3} D_{4,2} = \dfrac{1}{3} D_{2,4} = C_3^{(0)} = 0.164228.
\end{align}

\bibliographystyle{JHEP}
\bibliography{ref}

\providecommand{\href}[2]{#2}\begingroup\raggedright\begin{thebibliography}{10}

\bibitem{Friedberg:1976me}
R.~Friedberg, T.~D. Lee and A.~Sirlin, \emph{{A Class of Scalar-Field Soliton
  Solutions in Three Space Dimensions}},
  \href{https://doi.org/10.1103/PhysRevD.13.2739}{\emph{Phys. Rev. D}
  {\bfseries 13} (1976) 2739--2761}.

\bibitem{Coleman:1985ki}
S.~R. Coleman, \emph{{Q-balls}},
  \href{https://doi.org/10.1016/0550-3213(86)90520-1}{\emph{Nucl. Phys. B}
  {\bfseries 262} (1985) 263}.

\bibitem{Kusenko:1997si}
A.~Kusenko and M.~E. Shaposhnikov, \emph{{Supersymmetric Q balls as dark
  matter}}, \href{https://doi.org/10.1016/S0370-2693(97)01375-0}{\emph{Phys.
  Lett. B} {\bfseries 418} (1998) 46--54},
  [\href{https://arxiv.org/abs/hep-ph/9709492}{{\ttfamily hep-ph/9709492}}].

\bibitem{Dine:2003ax}
M.~Dine and A.~Kusenko, \emph{{The Origin of the matter - antimatter
  asymmetry}}, \href{https://doi.org/10.1103/RevModPhys.76.1}{\emph{Rev. Mod.
  Phys.} {\bfseries 76} (2003) 1},
  [\href{https://arxiv.org/abs/hep-ph/0303065}{{\ttfamily hep-ph/0303065}}].

\bibitem{Enqvist:2003gh}
K.~Enqvist and A.~Mazumdar, \emph{{Cosmological consequences of MSSM flat
  directions}},
  \href{https://doi.org/10.1016/S0370-1573(03)00119-4}{\emph{Phys. Rept.}
  {\bfseries 380} (2003) 99--234},
  [\href{https://arxiv.org/abs/hep-ph/0209244}{{\ttfamily hep-ph/0209244}}].

\bibitem{Cardoso:2019rvt}
V.~Cardoso and P.~Pani, \emph{{Testing the nature of dark compact objects: a
  status report}},
  \href{https://doi.org/10.1007/s41114-019-0020-4}{\emph{Living Rev. Rel.}
  {\bfseries 22} (2019) 4}, [\href{https://arxiv.org/abs/1904.05363}{{\ttfamily
  1904.05363}}].

\bibitem{PhysRev.172.1331}
D.~J. Kaup, \emph{Klein-gordon geon},
  \href{https://doi.org/10.1103/PhysRev.172.1331}{\emph{Phys. Rev.} {\bfseries
  172} (Aug, 1968) 1331--1342}.

\bibitem{PhysRev.187.1767}
R.~Ruffini and S.~Bonazzola, \emph{Systems of self-gravitating particles in
  general relativity and the concept of an equation of state},
  \href{https://doi.org/10.1103/PhysRev.187.1767}{\emph{Phys. Rev.} {\bfseries
  187} (Nov, 1969) 1767--1783}.

\bibitem{Liebling:2012fv}
S.~L. Liebling and C.~Palenzuela, \emph{{Dynamical boson stars}},
  \href{https://doi.org/10.1007/s41114-023-00043-4}{\emph{Living Rev. Rel.}
  {\bfseries 26} (2023) 1}, [\href{https://arxiv.org/abs/1202.5809}{{\ttfamily
  1202.5809}}].

\bibitem{Enqvist:2003zb}
K.~Enqvist and M.~Laine, \emph{{Q-ball dynamics from atomic Bose-Einstein
  condensates}},
  \href{https://doi.org/10.1088/1475-7516/2003/08/003}{\emph{JCAP} {\bfseries
  08} (2003) 003}, [\href{https://arxiv.org/abs/cond-mat/0304355}{{\ttfamily
  cond-mat/0304355}}].

\bibitem{Bunkov:2007fe}
Y.~M. Bunkov and G.~E. Volovik, \emph{{Magnons condensation into Q-ball in He-3
  - B}}, \href{https://doi.org/10.1103/PhysRevLett.98.265302}{\emph{Phys. Rev.
  Lett.} {\bfseries 98} (2007) 265302},
  [\href{https://arxiv.org/abs/cond-mat/0703183}{{\ttfamily
  cond-mat/0703183}}].

\bibitem{Battye:2000qj}
R.~Battye and P.~Sutcliffe, \emph{{Q-ball dynamics}},
  \href{https://doi.org/10.1016/S0550-3213(00)00506-X}{\emph{Nucl. Phys. B}
  {\bfseries 590} (2000) 329--363},
  [\href{https://arxiv.org/abs/hep-th/0003252}{{\ttfamily hep-th/0003252}}].

\bibitem{Bowcock}
P.~Bowcock, D.~Foster and P.~Sutcliffe, \emph{Q-balls, integrability and
  duality}, \href{https://doi.org/10.1088/1751-8113/42/8/085403}{\emph{Journal
  of Physics A: Mathematical and Theoretical} {\bfseries 42} (2009) 085403}.

\bibitem{Axenides:1999hs}
M.~Axenides, S.~Komineas, L.~Perivolaropoulos and M.~Floratos, \emph{{Dynamics
  of nontopological solitons: Q balls}},
  \href{https://doi.org/10.1103/PhysRevD.61.085006}{\emph{Phys. Rev. D}
  {\bfseries 61} (2000) 085006},
  [\href{https://arxiv.org/abs/hep-ph/9910388}{{\ttfamily hep-ph/9910388}}].

\bibitem{Copeland:2014qra}
E.~J. Copeland, P.~M. Saffin and S.-Y. Zhou, \emph{{Charge-Swapping Q-balls}},
  \href{https://doi.org/10.1103/PhysRevLett.113.231603}{\emph{Phys. Rev. Lett.}
  {\bfseries 113} (2014) 231603},
  [\href{https://arxiv.org/abs/1409.3232}{{\ttfamily 1409.3232}}].

\bibitem{Xie:2021glp}
Q.-X. Xie, P.~M. Saffin and S.-Y. Zhou, \emph{{Charge-Swapping Q-balls and
  Their Lifetimes}}, \href{https://doi.org/10.1007/JHEP07(2021)062}{\emph{JHEP}
  {\bfseries 07} (2021) 062},
  [\href{https://arxiv.org/abs/2101.06988}{{\ttfamily 2101.06988}}].

\bibitem{Zhou:2024mea}
S.-Y. Zhou, \emph{{Non-topological solitons and quasi-solitons}},
  \href{https://arxiv.org/abs/2411.16604}{{\ttfamily 2411.16604}}.

\bibitem{Bogolyubsky:1976nx}
I.~L. Bogolyubsky and V.~G. Makhankov, \emph{{On the Pulsed Soliton Lifetime in
  Two Classical Relativistic Theory Models}}, {\emph{JETP Lett.} {\bfseries 24}
  (1976) 12}.

\bibitem{Gleiser:1993pt}
M.~Gleiser, \emph{{Pseudostable bubbles}},
  \href{https://doi.org/10.1103/PhysRevD.49.2978}{\emph{Phys. Rev. D}
  {\bfseries 49} (1994) 2978--2981},
  [\href{https://arxiv.org/abs/hep-ph/9308279}{{\ttfamily hep-ph/9308279}}].

\bibitem{Copeland:1995fq}
E.~J. Copeland, M.~Gleiser and H.~R. Muller, \emph{{Oscillons: Resonant
  configurations during bubble collapse}},
  \href{https://doi.org/10.1103/PhysRevD.52.1920}{\emph{Phys. Rev. D}
  {\bfseries 52} (1995) 1920--1933},
  [\href{https://arxiv.org/abs/hep-ph/9503217}{{\ttfamily hep-ph/9503217}}].

\bibitem{Fodor:2008du}
G.~Fodor, P.~Forgacs, Z.~Horvath and M.~Mezei, \emph{{Computation of the
  radiation amplitude of oscillons}},
  \href{https://doi.org/10.1103/PhysRevD.79.065002}{\emph{Phys. Rev. D}
  {\bfseries 79} (2009) 065002},
  [\href{https://arxiv.org/abs/0812.1919}{{\ttfamily 0812.1919}}].

\bibitem{Fodor:2009kf}
G.~Fodor, P.~Forgacs, Z.~Horvath and M.~Mezei, \emph{{Radiation of scalar
  oscillons in 2 and 3 dimensions}},
  \href{https://doi.org/10.1016/j.physletb.2009.03.054}{\emph{Phys. Lett. B}
  {\bfseries 674} (2009) 319--324},
  [\href{https://arxiv.org/abs/0903.0953}{{\ttfamily 0903.0953}}].

\bibitem{Graham:2006xs}
N.~Graham and N.~Stamatopoulos, \emph{{Unnatural Oscillon Lifetimes in an
  Expanding Background}},
  \href{https://doi.org/10.1016/j.physletb.2006.06.070}{\emph{Phys. Lett. B}
  {\bfseries 639} (2006) 541--545},
  [\href{https://arxiv.org/abs/hep-th/0604134}{{\ttfamily hep-th/0604134}}].

\bibitem{Salmi:2012ta}
P.~Salmi and M.~Hindmarsh, \emph{{Radiation and Relaxation of Oscillons}},
  \href{https://doi.org/10.1103/PhysRevD.85.085033}{\emph{Phys. Rev. D}
  {\bfseries 85} (2012) 085033},
  [\href{https://arxiv.org/abs/1201.1934}{{\ttfamily 1201.1934}}].

\bibitem{Zhang:2020bec}
H.-Y. Zhang, M.~A. Amin, E.~J. Copeland, P.~M. Saffin and K.~D. Lozanov,
  \emph{{Classical Decay Rates of Oscillons}},
  \href{https://doi.org/10.1088/1475-7516/2020/07/055}{\emph{JCAP} {\bfseries
  07} (2020) 055}, [\href{https://arxiv.org/abs/2004.01202}{{\ttfamily
  2004.01202}}].

\bibitem{Olle:2020qqy}
J.~Olle, O.~Pujolas and F.~Rompineve, \emph{{Recipes for oscillon longevity}},
  \href{https://doi.org/10.1088/1475-7516/2021/09/015}{\emph{JCAP} {\bfseries
  09} (2021) 015}, [\href{https://arxiv.org/abs/2012.13409}{{\ttfamily
  2012.13409}}].

\bibitem{vanDissel:2023zva}
F.~van Dissel, O.~Pujolas and E.~I. Sfakianakis, \emph{{Oscillon
  spectroscopy}}, \href{https://doi.org/10.1007/JHEP07(2023)194}{\emph{JHEP}
  {\bfseries 07} (2023) 194},
  [\href{https://arxiv.org/abs/2303.16072}{{\ttfamily 2303.16072}}].

\bibitem{Gleiser:2011xj}
M.~Gleiser, N.~Graham and N.~Stamatopoulos, \emph{{Generation of Coherent
  Structures After Cosmic Inflation}},
  \href{https://doi.org/10.1103/PhysRevD.83.096010}{\emph{Phys. Rev. D}
  {\bfseries 83} (2011) 096010},
  [\href{https://arxiv.org/abs/1103.1911}{{\ttfamily 1103.1911}}].

\bibitem{Amin:2011hj}
M.~A. Amin, R.~Easther, H.~Finkel, R.~Flauger and M.~P. Hertzberg,
  \emph{{Oscillons After Inflation}},
  \href{https://doi.org/10.1103/PhysRevLett.108.241302}{\emph{Phys. Rev. Lett.}
  {\bfseries 108} (2012) 241302},
  [\href{https://arxiv.org/abs/1106.3335}{{\ttfamily 1106.3335}}].

\bibitem{Zhou:2013tsa}
S.-Y. Zhou, E.~J. Copeland, R.~Easther, H.~Finkel, Z.-G. Mou and P.~M. Saffin,
  \emph{{Gravitational Waves from Oscillon Preheating}},
  \href{https://doi.org/10.1007/JHEP10(2013)026}{\emph{JHEP} {\bfseries 10}
  (2013) 026}, [\href{https://arxiv.org/abs/1304.6094}{{\ttfamily 1304.6094}}].

\bibitem{Olle:2019kbo}
J.~Oll\'e, O.~Pujol\`as and F.~Rompineve, \emph{{Oscillons and Dark Matter}},
  \href{https://doi.org/10.1088/1475-7516/2020/02/006}{\emph{JCAP} {\bfseries
  02} (2020) 006}, [\href{https://arxiv.org/abs/1906.06352}{{\ttfamily
  1906.06352}}].

\bibitem{Aurrekoetxea:2023jwd}
J.~C. Aurrekoetxea, K.~Clough and F.~Muia, \emph{{Oscillon formation during
  inflationary preheating with general relativity}},
  \href{https://doi.org/10.1103/PhysRevD.108.023501}{\emph{Phys. Rev. D}
  {\bfseries 108} (2023) 023501},
  [\href{https://arxiv.org/abs/2304.01673}{{\ttfamily 2304.01673}}].

\bibitem{Ciurla:2024ksm}
D.~Ciurla, P.~Dorey, T.~Roma\'nczukiewicz and Y.~Shnir, \emph{{Perturbations of
  Q-balls: from spectral structure to radiation pressure}},
  \href{https://doi.org/10.1007/JHEP07(2024)196}{\emph{JHEP} {\bfseries 07}
  (2024) 196}, [\href{https://arxiv.org/abs/2405.06591}{{\ttfamily
  2405.06591}}].

\bibitem{Blaschke:2024dlt}
F.~Blaschke, T.~Roma\'nczukiewicz, K.~S\l{}awi\'nska and A.~Wereszczy\'nski,
  \emph{{Oscillons from Q-balls through Renormalization}},
  \href{https://doi.org/10.1103/PhysRevLett.134.081601}{\emph{Phys. Rev. Lett.}
  {\bfseries 134} (2025) 081601},
  [\href{https://arxiv.org/abs/2410.24109}{{\ttfamily 2410.24109}}].

\bibitem{Blaschke:2025anm}
F.~Blaschke, T.~Romanczukiewicz, K.~Slawinska and A.~Wereszczynski,
  \emph{{Q-ball polarization -- a smooth path to oscillons}},
  \href{https://arxiv.org/abs/2502.20519}{{\ttfamily 2502.20519}}.

\bibitem{Campbell:1983xu}
D.~K. Campbell, J.~F. Schonfeld and C.~A. Wingate, \emph{{Resonance structure
  in kink-antikink interactions in \ensuremath{\varphi} 4 theory }},
  \href{https://doi.org/10.1016/0167-2789(83)90289-0}{\emph{Physica D}
  {\bfseries 9} (1983) 1}.

\bibitem{Sugiyama:1979mi}
T.~Sugiyama, \emph{{Kink-Antikink collisions in the two-dimensional phi**4
  model}}, \href{https://doi.org/10.1143/PTP.61.1550}{\emph{Prog. Theor. Phys.}
  {\bfseries 61} (1979) 1550--1563}.

\bibitem{Manton:2021ipk}
N.~S. Manton, K.~Oles, T.~Romanczukiewicz and A.~Wereszczynski,
  \emph{{Collective Coordinate Model of Kink-Antikink Collisions in
  \ensuremath{\phi}4 Theory}},
  \href{https://doi.org/10.1103/PhysRevLett.127.071601}{\emph{Phys. Rev. Lett.}
  {\bfseries 127} (2021) 071601},
  [\href{https://arxiv.org/abs/2106.05153}{{\ttfamily 2106.05153}}].

\bibitem{Takyi:2016tnc}
I.~Takyi and H.~Weigel, \emph{{Collective Coordinates in One-Dimensional
  Soliton Models Revisited}},
  \href{https://doi.org/10.1103/PhysRevD.94.085008}{\emph{Phys. Rev. D}
  {\bfseries 94} (2016) 085008},
  [\href{https://arxiv.org/abs/1609.06833}{{\ttfamily 1609.06833}}].

\bibitem{Lima:2021jxl}
F.~C. Lima, F.~C. Simas, K.~Z. Nobrega and A.~R. Gomes, \emph{{Scattering of
  metastable lumps in a model with a false vacuum}},
  \href{https://doi.org/10.1016/j.physletb.2021.136707}{\emph{Phys. Lett. B}
  {\bfseries 822} (2021) 136707},
  [\href{https://arxiv.org/abs/2108.13579}{{\ttfamily 2108.13579}}].

\bibitem{Campos:2021mkn}
J.~a. G.~F. Campos and A.~Mohammadi, \emph{{Wobbling double sine-Gordon
  kinks}}, \href{https://doi.org/10.1007/JHEP09(2021)067}{\emph{JHEP}
  {\bfseries 09} (2021) 067},
  [\href{https://arxiv.org/abs/2103.04908}{{\ttfamily 2103.04908}}].

\bibitem{Krusch:2024vuy}
S.~Krusch, M.~Rees and T.~Winyard, \emph{{Scattering of vortices with excited
  normal modes}},
  \href{https://doi.org/10.1103/PhysRevD.110.056050}{\emph{Phys. Rev. D}
  {\bfseries 110} (2024) 056050},
  [\href{https://arxiv.org/abs/2406.04164}{{\ttfamily 2406.04164}}].

\bibitem{AlonsoIzquierdo:2024nbn}
A.~Alonso~Izquierdo, N.~S. Manton, J.~Mateos~Guilarte and A.~Wereszczynski,
  \emph{{Collective coordinate models for 2-vortex shape mode dynamics}},
  \href{https://doi.org/10.1103/PhysRevD.110.085006}{\emph{Phys. Rev. D}
  {\bfseries 110} (2024) 085006},
  [\href{https://arxiv.org/abs/2405.20249}{{\ttfamily 2405.20249}}].

\bibitem{Blaschke:2024uec}
F.~Blaschke, T.~Roma\'nczukiewicz, K.~S\l{}awi\'nska and A.~Wereszczy\'nski,
  \emph{{Amplitude modulations and resonant decay of excited oscillons}},
  \href{https://doi.org/10.1103/PhysRevE.110.014203}{\emph{Phys. Rev. E}
  {\bfseries 110} (2024) 014203},
  [\href{https://arxiv.org/abs/2403.00443}{{\ttfamily 2403.00443}}].

\bibitem{Manton:1981mp}
N.~S. Manton, \emph{{A Remark on the Scattering of BPS Monopoles}},
  \href{https://doi.org/10.1016/0370-2693(82)90950-9}{\emph{Phys. Lett. B}
  {\bfseries 110} (1982) 54--56}.

\bibitem{Blaschke:2025qkg}
F.~Blaschke, T.~Roma\'nczukiewicz, K.~S\l{}awi\'nska and A.~Wereszczy\'nski,
  \emph{{Oscillons from Q-balls}},
  \href{https://doi.org/10.1103/PhysRevD.111.036034}{\emph{Phys. Rev. D}
  {\bfseries 111} (2025) 036034},
  [\href{https://arxiv.org/abs/2502.09136}{{\ttfamily 2502.09136}}].

\bibitem{Manton:2023mdr}
N.~S. Manton and T.~Roma\'nczukiewicz, \emph{{Simplest oscillon and its
  sphaleron}}, \href{https://doi.org/10.1103/PhysRevD.107.085012}{\emph{Phys.
  Rev. D} {\bfseries 107} (2023) 085012},
  [\href{https://arxiv.org/abs/2301.09660}{{\ttfamily 2301.09660}}].

\bibitem{Navarro-Obregon:2023hqe}
S.~Navarro-Obreg\'on, L.~M. Nieto and J.~M. Queiruga, \emph{{Inclusion of
  radiation in the collective coordinate method approach of the
  \ensuremath{\phi}4 model}},
  \href{https://doi.org/10.1103/PhysRevE.108.044216}{\emph{Phys. Rev. E}
  {\bfseries 108} (2023) 044216},
  [\href{https://arxiv.org/abs/2305.00497}{{\ttfamily 2305.00497}}].

\bibitem{Adam:2021gat}
C.~Adam, N.~S. Manton, K.~Oles, T.~Romanczukiewicz and A.~Wereszczynski,
  \emph{{Relativistic moduli space for kink collisions}},
  \href{https://doi.org/10.1103/PhysRevD.105.065012}{\emph{Phys. Rev. D}
  {\bfseries 105} (2022) 065012},
  [\href{https://arxiv.org/abs/2111.06790}{{\ttfamily 2111.06790}}].

\bibitem{GarciaMartin-Caro:2025zkc}
A.~Garc\'\i{}a Mart\'\i{}n-Caro, J.~Queiruga and A.~Wereszczynski,
  \emph{{Feshbach resonances and dynamics of BPS solitons}},
  \href{https://arxiv.org/abs/2501.02589}{{\ttfamily 2501.02589}}.

\end{thebibliography}\endgroup
\end{document}